\documentclass[fleqn,usenatbib]{mnras}

\usepackage{newtxtext,newtxmath}
\usepackage{bm}

\usepackage[T1]{fontenc}
\usepackage{ae,aecompl}

\usepackage{graphicx}	
\usepackage{amsmath}	
\usepackage{amssymb}	

\title[Polar alignment of a disc around a
  binary]{Polar alignment of a protoplanetary disc around an
  eccentric binary II: Effect of binary and disc parameters}

\author[Martin \& Lubow]{Rebecca G. Martin$^{1}$ and Stephen
  H. Lubow$^2$\\ 
$^{1}$Department of Physics and Astronomy, University
  of Nevada, Las Vegas, 4505 South Maryland Parkway, Las Vegas, NV
  89154, USA \\ 
$^2$Space Telescope Science Institute, 3700 San Martin
  Drive, Baltimore, MD 21218, USA}

\date{Accepted XXX. Received YYY; in original form ZZZ}

\pubyear{2018}

\begin{document}

\label{firstpage}
\pagerange{\pageref{firstpage}--\pageref{lastpage}}
\maketitle


\begin{abstract}
In a recent paper Martin \& Lubow showed that a circumbinary disc around an eccentric binary can undergo damped nodal oscillations that lead to the polar (perpendicular) alignment of the disc relative to the binary orbit. The disc angular momentum vector aligns to the eccentricity vector of the binary. We explore the robustness of this mechanism for a low mass disc (0.001 of the binary mass) and its dependence on system parameters by means of hydrodynamic disc simulations. We describe how the evolution depends upon the disc viscosity, temperature, size,  binary mass ratio, orbital eccentricity and inclination. We compare results with predictions of linear theory.
We show that polar alignment of a low mass disc may occur over a wide range of binary--disc parameters. We discuss the application of our results to the formation of planetary systems around eccentric binary stars.

\end{abstract}

\begin{keywords} accretion, accretion discs -- binaries: general --
  hydrodynamics -- planets and satellites: formation
\end{keywords}



\section{Introduction}
\label{intro}

Binary and multiple star systems are observed to be common
\citep{Ghez1993,Duchene2013}. Turbulent molecular clouds can lead to
chaotic accretion during the star formation process
\citep{McKee2007,Bate2003} and thus circumstellar and circumbinary
discs are likely to form misaligned with respect to the orbital plane of the binary
\citep{Monin2007,Bateetal2010,Bate2018}. The torque from the binary system alters
the formation processes for planets in these discs compared to discs
around single stars \citep{Nelson2000,Mayer2005,Boss2006, Martinetal2014b, Fu2017}. After a giant
planet forms in a misaligned disc, the torque from the binary can
result in the planet no longer remaining coplanar to the disc
\citep{Picogna2015,Lubow2016,Martin2016}. Understanding the disc evolution in
misaligned systems is critical to explaining properties of exoplanets.

Observations of circumbinary discs have revealed misalignments between
the disc and the orbital plane of the binary. For example, the
peculiar light curve of KH 15D can be explained by a misaligned
precessing disc \citep[e.g.,][]{Winn2004, Chiang2004,Capelo2012}. The
binary protostar IRS~43 has a misalignment greater than $60^\circ$
between the binary and the disc \citep{Brinch2016}. The debris disc
around 99 Herculis is most likely perpendicular to the binary
eccentricity vector \citep{Kennedy2012}.

A misaligned circular test particle orbit around a {\it circular} orbit binary undergoes uniform
nodal precession with constant tilt. The angular momentum vector of the test particle
precesses about the binary angular momentum vector. Similarly, a misaligned
disc around a circular binary displays the form of same precession. The disc
is able to hold itself together and precess as a solid body for
sufficiently warm and narrow discs
\citep[e.g.,][]{LP1997}. Dissipation within the disc leads to
alignment with the binary orbital plane
\citep{PT1995, Lubow2000, Nixonetal2011b,Nixon2012,Facchinietal2013,
  Lodato2013,Foucart2013,Foucart2014}.

However, main sequence binaries are commonly found on eccentric orbits \citep[e.g.,][]{Duquennoy1991, Raghavan2010}.
The eccentricity distribution is nearly flat out to $e \sim 0.6$
and declines somewhat at higher values.
 Consequently, the binary eccentricity should
be considered  in dynamical models of interactions between a stellar binary and a disc.
  Misaligned test particle orbits around an eccentric orbit binary generally undergo
nonuniform precession with time varying tilt \citep{Verrier2009,Farago2010,Doolin2011}. 

For sufficiently high inclination, these orbits 
can precess around the binary eccentricity vector, rather than the binary angular momentum vector.
Such particles undergo a
nodal libration process.
Oscillations occur between the tilt of the particle orbit and the
longitude of the ascending node.
 The angular momentum vector of the test particle precesses about the
eccentricity vector of the binary, while the longitude of the ascending node librates. There is a critical initial inclination above which the
mechanism operates. Below this critical inclination, the test particle
angular momentum vector precesses about the binary angular momentum
vector, while the longitude of the ascending node circulates. The critical angle
depends upon the binary eccentricity.

Recently we showed that, for typical parameters of protoplanetary
circumbinary discs, a moderately inclined, low mass circumbinary disc
around an eccentric orbit binary can also precess about the binary
eccentricity vector \citep[see also][]{Aly2015}. Furthermore, unlike
the test particle case, the disc evolves to a state in which the disc
lies perpendicular to the binary orbital plane \citep{Martin2017}.
Linear models for the evolution of a low mass, inclined disc around an
eccentric binary also show that a circumbinary disc can evolve to a
perpendicular state in less than the disc lifetime for a plausible
range of binary and disc parameters \citep[][hereafter
  LM18]{Zanazzi2018,Lubow2018}.  As in the particle case, the disc is
subject to libration of the tilt angle and the longitude of the
ascending node. Dissipation within the disc causes the oscillations to
damp. The disc evolves to polar alignment in which the angular
momentum vector of the disc aligns with the binary eccentricity
vector.  The disc then settles to a state in which it lies
perpendicular to the orbital plane of the binary.

 In this work we examine this process in more detail for a low mass disc by exploring the
parameter space of binary and disc properties. 
In this regime, the mass and angular momentum of the disc are sufficiently small
that the effects of the disc on the binary evolution are also small. We note that the case 
of an extreme mass ratio binary was studied by \cite{Teyssandier2013} who considered the interaction of a planet orbiting a star with an external disc around both the star and planet. In this case, the angular momentum of the disc is large compared to that of the star-planet system.
The planet undergoes Kozai--Lidov oscillations \citep{Kozai1962,Lidov1962} for a sufficiently inclined disc. In this work we consider binary mass ratios greater than $0.1$, relevant to binary star systems, that result in quite different behaviour.  

In Section~\ref{disc} we describe hydrodynamic simulations to examine the effect of disc properties such as the viscosity, the temperature, the size and the mass, and binary properties such as the inclination, mass ratio, and eccentricity. We compare some of the results of simulations to linear theory
in Section~\ref{linth}. We discuss the implications of our results for planet formation around binary stars in Section~\ref{discussion} and draw our conclusions in Section~\ref{conc}.

\section{Circumbinary disc simulations}
\label{disc}

We explore the evolution of a misaligned circumbinary disc around a
binary star system with the smoothed particle hydrodynamics (SPH;
e.g. \citealt{Price2012a}) code {\sc phantom}
\citep{PF2010,LP2010,Price2017}. {\sc phantom} has been used extensively for
simulations of misaligned accretion discs
\citep[e.g.][]{Nixon2012,Nixonetal2013,Martinetal2014b,Fu2015}. Table~\ref{tab}
shows a list of simulation parameters for simulations that we
describe in detail in this section. The binary has total mass
$M=M_1+M_2$ and orbits with semi--major axis $a_{\rm b}$ and eccentricity $e_{\rm b}$. The origin is at the binary centre of mass and the distance from the centre of mass is denoted by $R$.   
The initial orbit of the binary lies in the Cartesian $x-y$ plane.
 The $y$--axis  is defined such that it lies along the direction of the initial binary eccentricity vector. The binary components initially reside at apastron.

The circumbinary disc is initially flat and tilted with
respect to the binary orbital plane by inclination $i$.  In this work
we ignore effects of self-gravity and consider only discs with small initial disc mass $M_{\rm d}=0.001\, M$.
The orbital elements of the disc that we describe, such as inclination $i$ and longitude of ascending node $\phi$,
are defined relative to the initial orbit of the binary.  
That is, the reference plane for the orbital elements is the initial binary orbital plane
and the reference direction is along the initial binary eccentricity vector.
Due to the nonzero disc mass, the binary orbit is not
fixed in space, but the changes in the orbit are typically small for a disc of this mass. For example, for the fiducial disc model in Section 2.1, the binary semi--major axis changes by less than 1\% and the eccentricity by less than 4\% during a time of $1000$ binary orbits. We consider some larger disc radii in Section 2.4. For the largest we consider of $R_{\rm out}=50\,a_{\rm b}$, the change to the binary semi--major axis is less than 4\% while the eccentricity changes by less than 16\%.

 The initial disc surface density is taken to follow a power law distribution $\Sigma \propto R^{-3/2}$
between initial inner radius $R_{\rm in}=2\,a_{\rm b}$ out to some initial
outer radius $R_{\rm out}$, which we vary between models. The disc can evolve across these
radii. The initial
inner disc truncation radius is chosen to be that of a tidally
truncated coplanar disc \citep{Artymowicz1994}. We note that the tidal
torque on a misaligned disc is weaker and so the disc may extend closer
to the binary in this case
\citep[e.g.][LM18]{Lubowetal2015,Nixon2015,Miranda2015}. The disc is
locally isothermal with sound speed $c_{\rm s}\propto R^{-3/4}$ and
therefore the disc aspect ratio is $H/R\propto R^{-1/4}$.  These dependences imply
that $\alpha$ and the smoothing length $\left<h \right>/H$ are
constant over the radial extent of the disc \citep{LP2007}.  We choose
the value of the disc aspect ratio $H/R$ at the initial inner disc edge, $R_{\rm
  in}=2\,a_{\rm b}$. We consider two different values of the \cite{SS1973} $\alpha$ viscosity parameter in our
simulations. The disc viscosity is implemented in the usual manner by
adapting the SPH artificial viscosity according to \cite{LP2010}.  For
our simulations with $3 \times 10^5$ particles and initial outer disc radius $R_{\rm out}=5\,a$, the disc is resolved with shell-averaged smoothing length per scale height $\left<h\right> /H \approx 0.25$.  Simulations with larger outer disc radius have a lower resolution.  For $R_{\rm out}=10\,a$, $\left<h\right> /H \approx 0.50$, for $R_{\rm out}=20\,a$, $\left<h\right> /H \approx 0.57$ initially and for $R_{\rm out}=50\,a$, $\left<h\right> /H \approx 0.69$. The viscosity parameter is held fixed between the different resolutions.

\begin{figure}
\centering
\includegraphics[width=8.0cm]{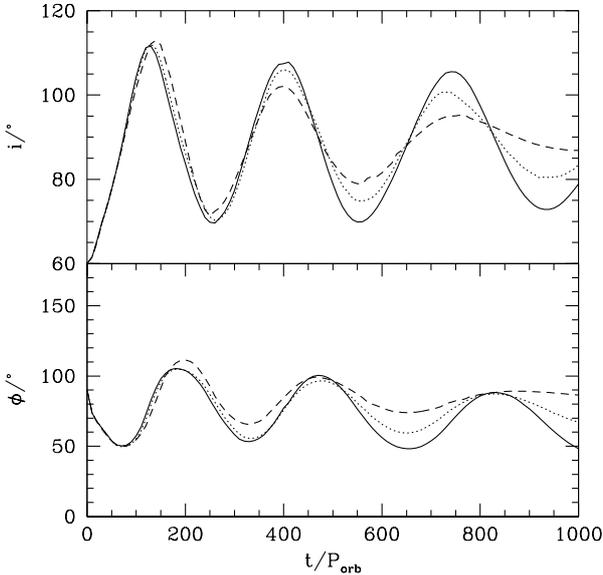}
\caption{Upper panel: The inclination of the angular momentum of the disc to that of the binary. Lower panel: The precession phase angle or longitude of ascending node. Standard disc parameters from \protect\cite{Martin2017} at a
  radius of $3\,a_{\rm b}$ for run1 in Table~\ref{tab} for three
  different resolutions. The solid line has $10^6$ initial particles, the
  dotted line has $3\times10^5$ initial particles, and the dashed line has
  $10^5$ initial particles. }
\label{paper1}
\end{figure}

\begin{table*}
\caption{Parameters of the initial circumbinary disc set up for binary
  with total mass $M$ and separation $a$. Disc aspect ratio $H/R$ is at $R=2 a_{\rm b}$. The disc may display a circulating (C) or librating (L) solution.}  \centering
\begin{tabular}{lllllllllllll}
\hline
Name & Fig.   & $M_1/M$ & $M_2/M$ & $e_{\rm b}$ &  $R_{\rm out}/a$  & $i/^\circ$& $H/R$ & $\alpha$ & C or L\\
\hline
\hline
run1 & \ref{paper1}  & 0.5 & 0.5 & 0.5& 5 &60 & 0.1 & 0.01 & {\rm L} \\
\hline
run2 & \ref{a} (left)   & 0.5 & 0.5 &0.5& 5  & 60& 0.1 &  0.1& {\rm L} \\
\hline
run3 & \ref{a} (right) &  0.5 & 0.5 & 0.5&5 &60 & 0.01 & 0.01& {\rm L} \\
\hline
run4 & \ref{size} (left)&   0.5 & 0.5 & 0.5&10 & 60 & 0.1 & 0.01& {\rm L} \\
run5 & \ref{size} (middle) & 0.5 & 0.5 & 0.5&20 & 60& 0.1 & 0.01 & {\rm L} \\
run6 & \ref{size} (right), \ref{warp}  & 0.5 & 0.5 &0.5& 50 & 60& 0.1 & 0.01& {\rm L}  \\
\hline
run7 & \ref{inc}  &0.5 & 0.5 &0.5& 5 &  20 & 0.1 &0.01& {\rm C} \\
run8 &   &0.5 & 0.5 &0.5& 5 &  40 & 0.1 &0.01& {\rm C} \\
run9 &   &0.5 & 0.5 &0.5& 5 &  50 & 0.1 &0.01& {\rm L } \\
run10 &\ref{inc} & 0.5 & 0.5 &0.5& 5 &  160 & 0.1 &0.01 & {\rm C}\\
\hline
run11 & \ref{q}   &   0.7 & 0.3 &0.5& 5 & 60& 0.1 & 0.01& {\rm L} \\
run12 & \ref{q}   &   0.9 & 0.1 &0.5& 5 & 60& 0.1 & 0.01& {\rm L} \\
\hline
run13 &\ref{kh15d}   &   0.5 & 0.5 &0.8& 5 & 10& 0.1 & 0.01& {\rm C} \\
run14 &\ref{kh15d}    &   0.5 & 0.5 & 0.8& 5 & 20& 0.1 & 0.01& {\rm C} \\
run15 &               & 0.5& 0.5& 0.8 &5&  30 & 0.1 & 0.01 & {\rm L}\\
run16 &\ref{kh15d}    &   0.5 & 0.5 & 0.8& 5 & 40& 0.1 & 0.01& {\rm L} \\
run17 &\ref{kh15d}    &   0.5 & 0.5 & 0.8& 5 & 60& 0.1 & 0.01 & {\rm L}\\
\hline
run18 & \ref{nearlypolar} (left)  & 0.5 & 0.5 & 0.5 & 5  & 80 & 0.1 & 0.01 & {\rm L} \\
run19 & \ref{nearlypolar} (right)  & 0.5 & 0.5 & 0.5 & 5  & 80 & 0.1 & 0.1 & {\rm L} \\
\hline
\end{tabular}
\label{tab}
\end{table*}

We analyse the data from the SPH simulations by dividing the disc
into 500 bins in spherical radius $R$. The bins extend from the initial disc inner radius to the initial disc outer radius.
Within each bin we calculate the
mean properties of the particles such as the surface density,
inclination and longitude of ascending node.  As the disc evolves, some disc material accretes towards the binary and the disc expands outward somewhat.
Particles are removed from the simulation when they reach within a radius $R_{\rm acc} = 0.25 a_{\rm b}$ about either
star. 

\subsection{Resolution}
\begin{figure}
\centering
\includegraphics[width=8cm]{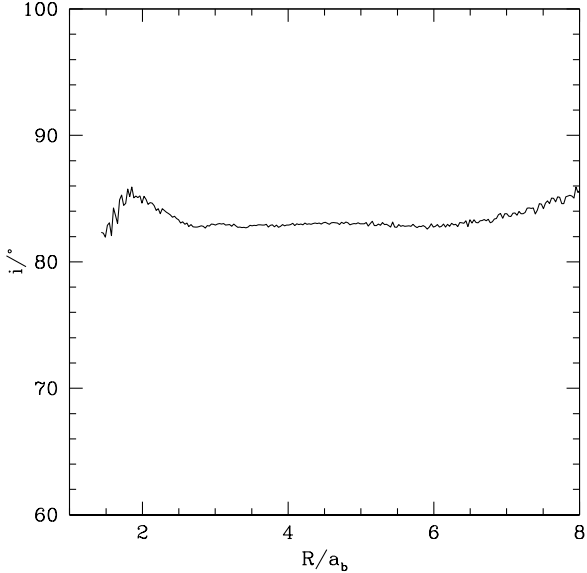}
\caption{The inclination of the disc as a function of radius for run1 at a time of $t=500\,P_{\rm orb}$ for the simulation with $3\times 10^5$ particles initially.}
\label{inc500}
\end{figure}

We first describe
 a resolution study. 
Fig.~\ref{paper1} shows the disc inclination and
longitude of the ascending node evolution for our standard parameters
taken from \cite{Martin2017}. These parameters are shown in run1 in
Table~\ref{tab}. The binary is equal mass with an orbital eccentricity
of 0.5.  The disc is initially flat and inclined by $60^\circ$ with respect to
the binary orbital plane.  The disc aspect ratio is $H/R=0.1$ at
$R=R_{\rm in}=2\,a_{\rm b}$.  The initial disc outer radius $R_{\rm out}$
 is $5\,a_{\rm b}$. The disc undergoes damped oscillations and
evolves towards polar alignment. We show three different resolutions,
$10^5$, $3\times 10^5$, and $10^6$ particles. In these simulations, over most of its radial extent, the
disc remains flat and we plot the evolution at a disc radius
of $R=3\,a_{\rm b}$.  Fig.~\ref{inc500} shows the inclination of the disc as a function of radius at a time of $t=500\,P_{\rm orb}$ for the disc that initially has $3\times 10^5$  particles.  Notice that the disc is quite flat between $R=2.5 a_{\rm b}$ and $R=7 a_{\rm b}$ which contains 86\% of the disc mass at that time.
    However, the inner and outer parts of the disc undergo substantial warping.

As described in \cite{Martin2017}, the disc
undergoes librations of the longitude of the ascending node and tilt. Dissipation causes the disc to evolve towards
polar alignment. That is, tilt $i$ undergoes damped oscillations that are centered about $i \simeq 90^{\circ}$.
 The disc angular momentum vector then evolves towards alignment with the
eccentricity vector of the binary. For a time of about $500\,P_{\rm
  orb}$, the results at the two highest resolutions are very similar. At later times, the 
  resolution degrades due to the reduction in the density of particles in the circumbinary disc. As a result,
  the evolution curves for the two highest resolutions begin to noticeably depart. In
order to present a large number of simulations in this work, we choose
to use $3\times10^5$ particles in all of the following simulations. We concentrate on times less than $\sim 500\,P_{\rm
  orb}$ for models with $\alpha=0.01$ and $H/r=0.1$ and times less than $\sim 50\,P_{\rm
  orb}$ for models with $\alpha=0.1$ and $H/r=0.1$ that are expected to evolve about ten times faster (see LM18).
  In some cases, we show plots that extend somewhat longer in time, where we examine more qualitative behaviour.  The resolution does not significantly affect the phasing of the tilt oscillations, only the amplitude. Low resolution in the simulations artificially enhances the viscosity. Thus, the amplitude of the oscillations may be underestimated on the longer timescales shown.

\subsection{Effect of disc viscosity}

\begin{figure}
\includegraphics[width=8cm]{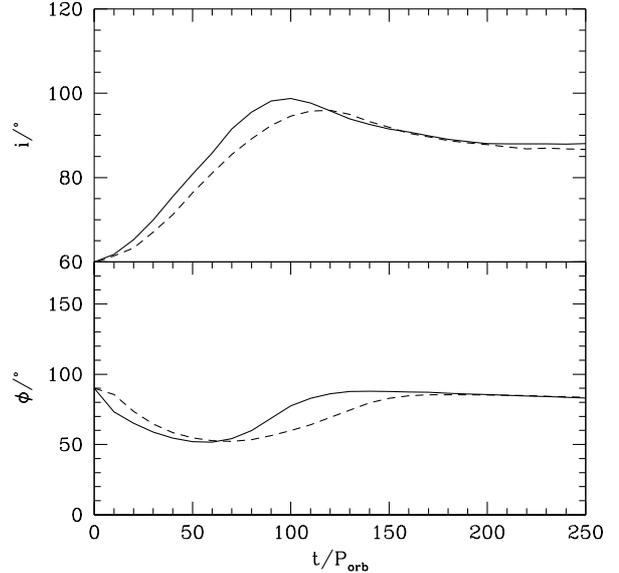}
\caption{Same as Fig.~\ref{paper1} with $3 \times 10^5$ particles, except that $\alpha=0.1$
  (run2). The solid line shows the evolution of the
  inner disc at a radius of $3\,a_{\rm b}$, while the dashed line shows
  the evolution of the outer disc at a radius of $5\,a_{\rm b}$. }
\label{a}
\end{figure}

We consider the evolution with our standard initial misalignment angle of $60^\circ$ as in Fig.~\ref{paper1}. Fig.~\ref{a} shows the disc evolution with a higher
viscosity parameter $\alpha=0.1$ for run2.  All other parameters are the same
as the standard parameters in run1 shown in Fig.~\ref{paper1}. The time for the first peak in the
inclination is not affected significantly compared to Fig.~\ref{paper1}. However, the higher
viscosity leads to stronger damping and the disc evolves towards polar
alignment more rapidly. There is only one oscillation during this
time. This result is not surprising because the tilt evolution rate  increases with the disc viscosity (LM18).
The figure also shows that the evolution at two different disc radii,
$3 a_{\rm b}$ and $5 a_{\rm b}$ is quite similar, indicating that the disc is quite
flat over these radii.

\subsection{Effect of the disc temperature}

\begin{figure*}
\includegraphics[width=8cm]{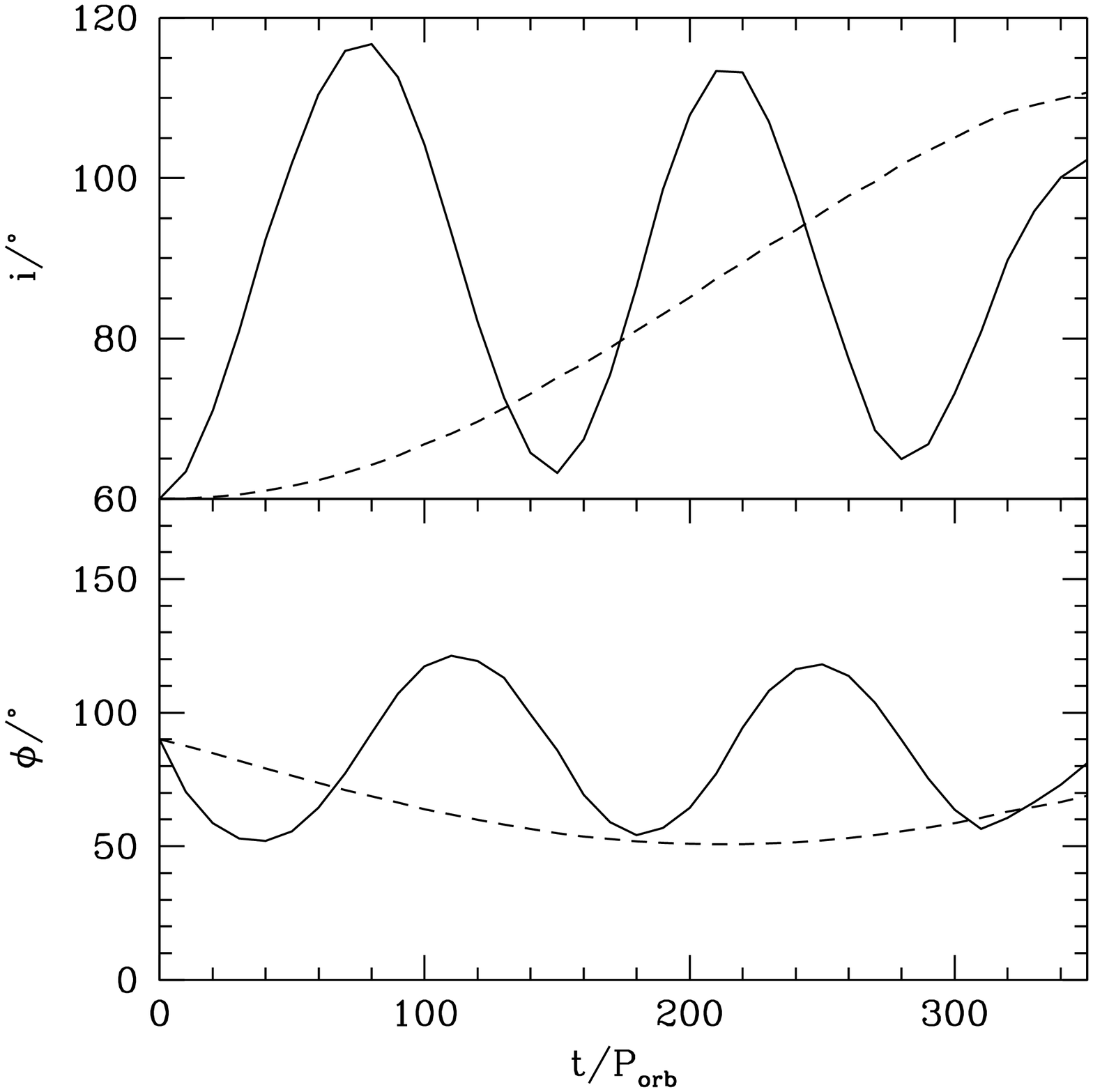}
\includegraphics[width=8cm]{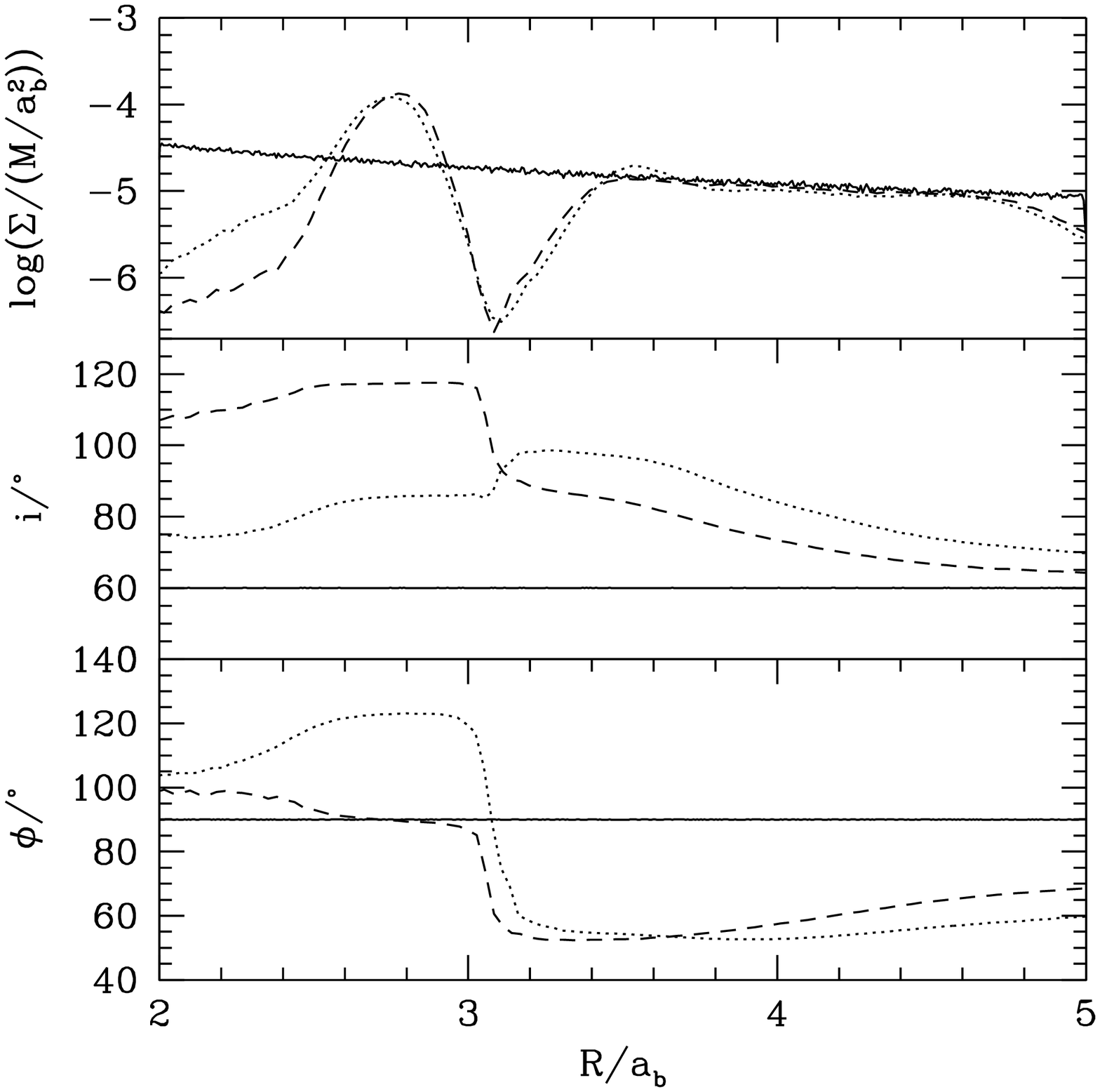}
\caption{Simulation with $H/R=0.01$ (run3). Left: Same as Fig.~\ref{paper1} with $3 \times 10^5$ particles. The solid line shows the evolution of the inner disc at a radius of $2.5\,a_{\rm b}$, while the dashed line shows  the evolution of the outer disc at a radius of $5\,a_{\rm b}$. Right: The surface density (upper panel), inclination (middle panel) and the phase angle (lower panel) at times $t=0$ (solid lines), $t=80$ (dashed lines) and $t=120\,P_{\rm orb}$ (dotted lines) as a function of the spherical radius from the centre of mass of the binary. }
\label{temp}
\end{figure*}

Fig.~\ref{temp} shows a disc with a lower disc aspect
ratio of $H/R=0.01$. The left panel shows the inclination and phase angle evolution at radii $R=2.5 a_{\rm b}$ and $R=5 a_{\rm b}$.  Such a cool disc is unable to hold itself together, if the disc precession timescale is shorter than the sound crossing time \citep[see also][LM18]{Aly2015}. The right panel shows the surface density, inclination and phase angle at three different times as a function of spherical radius from the center of mass of the binary.  This disc breaks into two disjoint discs with the break at a radius of around $3\,a_{\rm b}$. At later times more rings break off from the outer disc that is significantly warped. During the evolution  shown in Fig~\ref{temp} there are just two rings. This behaviour has already been seen in circumbinary disc simulations around circular binaries \citep{Nixonetal2012b,NK2012}. The inner and the outer parts of
the disc precess independently and show tilt oscillations on different
timescales. The cool disc is on the boundary of the viscous regime where
$\alpha>H/R$ and radial communication is diffusive. \cite{Aly2015}
investigated this regime in detail. The standard parameters that we use for
protoplanetary discs are in the wave--like regime with $\alpha<H/R$
and communication is a consequence of radial pressure forces.

To analyse these results further, we calculate the breaking radius $R_{\rm break}$, inside of which the disc
experiences strong warping that could result in disc breaking. We apply equation (34)
of LM18. This equation assumes a power law dependences of $\Sigma \propto 1/R^p$
and $T \propto 1/R^s$. In this paper $s=1.5$ at all times, while $p=1.5$ and $R_{\rm out}=5 a_{\rm b}$ initially.
Although the density evolves to a nonpower law distribution and $R_{\rm out}$ increases in time,  this density profile roughly holds
over much of the disc  and most of the disc mass remains within $R < 5 a_{\rm b}$ over the evolution times in the figure. 
With these parameters, we have that $R_{\rm break} \sim 1.7 a_{\rm b}$  for the 
case of $H/R=0.1$ and  $R_{\rm break} \sim 4.3 a_{\rm b}$ for the 
case of $H/R=0.01$.  The warmer disc is then not expected to experience breaking within its initial radial range ($R_{\rm in}=2 a_{\rm b}$),
while the cooler disc is expected to experience strong breaking over its main body as we see in the figure. 
The breaking radius estimate  is crude, but agrees with the results that that the cooler disc undergoes breaking in its main body,
while the warmer disc does not.

\subsection{Effect of the  disc size}
\label{sec:discsize}
\begin{figure*}
\centering \includegraphics[width=5.8cm]{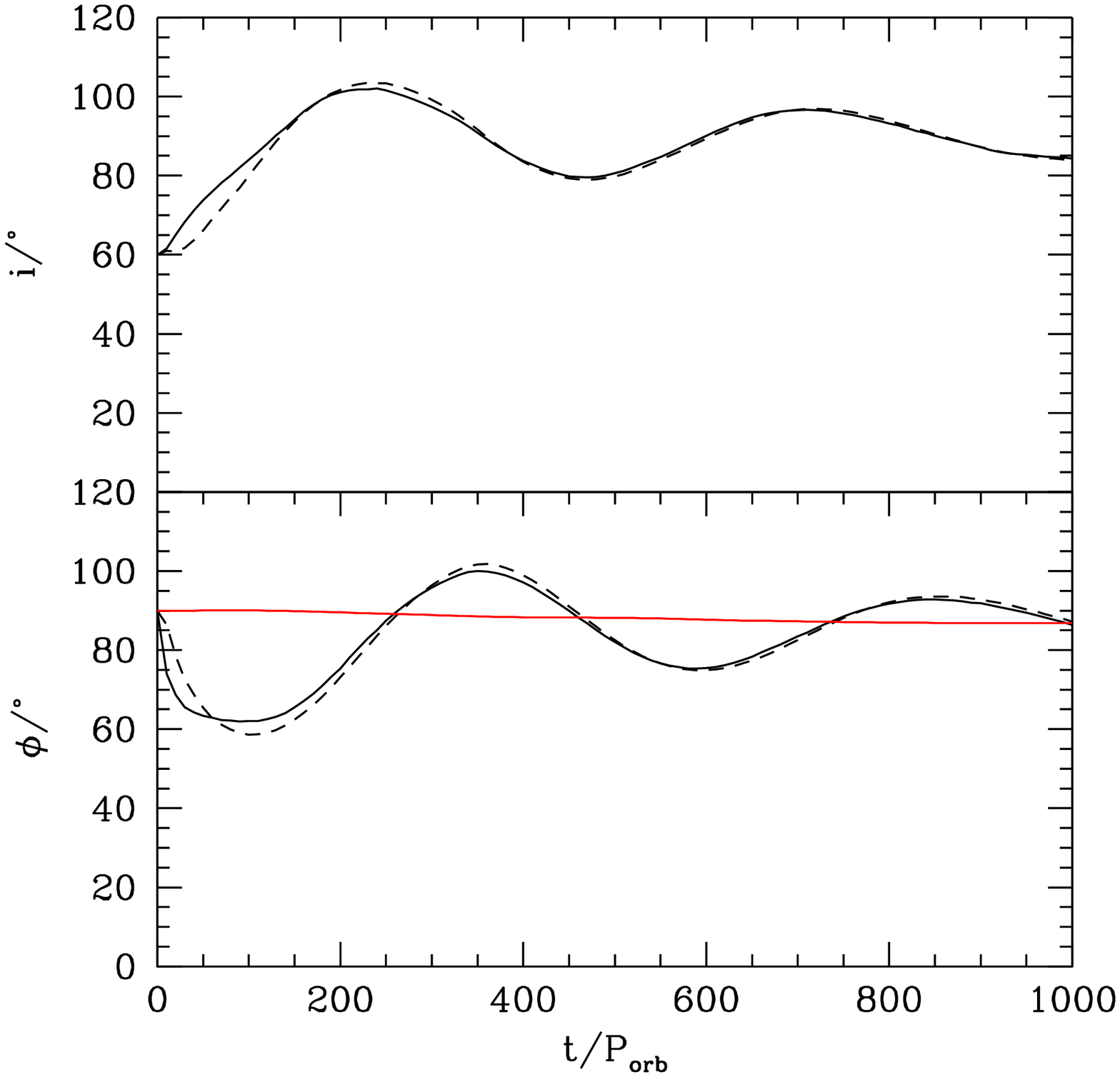}
\includegraphics[width=5.8cm]{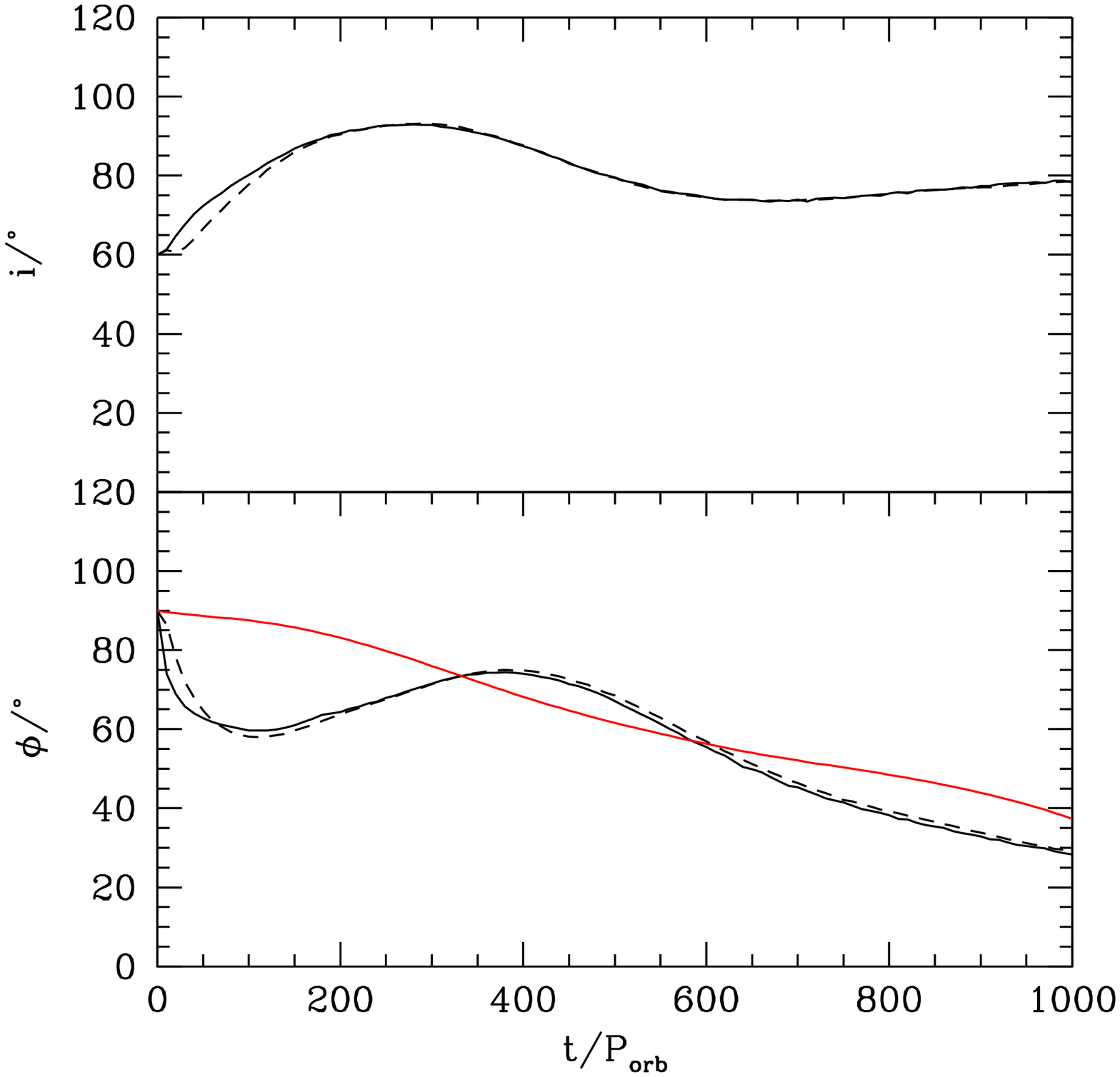}
\includegraphics[width=5.8cm]{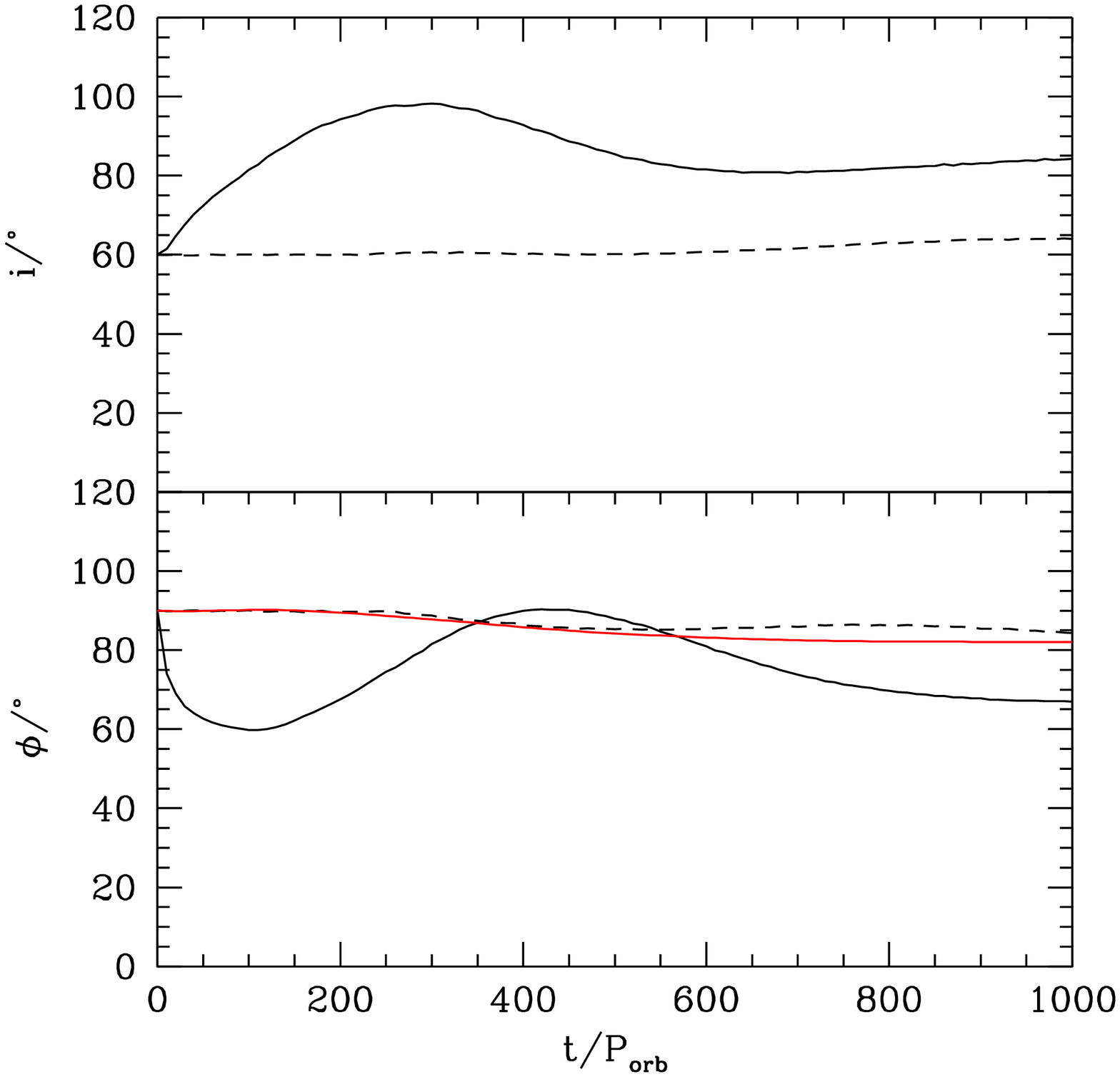}
\caption{Same as Fig.~\ref{paper1} with $3 \times 10^5$ particles, except for the initial outer 
  radius of the disc.  Left panel: $R_{\rm out}=10\,a_{\rm b}$ (run4) at
  radii $R=3 \,a_{\rm b}$ (solid line) and $R=10\, a_{\rm b}$ (dashed line). Middle:
  $R_{\rm out}=20\,a_{\rm b}$ (run5) at radii $R=3 \,a_{\rm b}$ (solid line) and
  $R=20\, a_{\rm b}$ (dashed line). Right: $R_{\rm out}=50\,a_{\rm b}$ (run6) at $R=3
  \,a_{\rm b}$ (solid line) and $R=50 \, a_{\rm b}$ (dashed line). The red lines plot
  the azimuthal angle of the eccentricity vector of the binary that is defined by Equation~(\ref{ephi}).  
  }
\label{size}
\end{figure*}

\begin{figure}
\centering
\includegraphics[width=8.0cm]{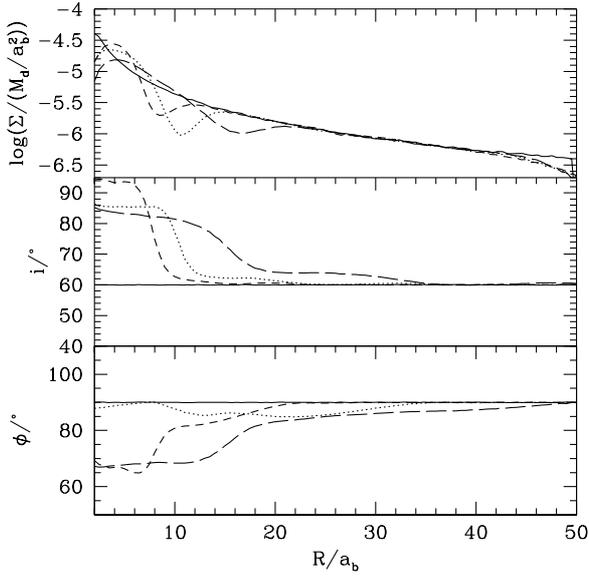}
\caption{The parameters are the same as Fig.~\ref{paper1} with $3 \times 10^5$ particles except the initial outer disc radius is $R_{\rm out}=50 \,a_{\rm   b}$ (run6). The surface density (upper panel), inclination  (middle panel) and precession angle (lower panel) are plotted as a function of  the spherical radius from the centre of mass of the binary at times  $0$ (solid lines), $250$ (short--dashed lines), $500$ (dotted  lines) and $1000\,P_{ \rm orb}$ (long--dashed lines). The plot shows the effect of a large initial outer radius of the disc on the disc structure.}
\label{warp}
\end{figure}

The size of a circumbinary disc relative to the binary separation can have a wide range of values. For example, the disc around KH~15D is thought to be a ring that extends to a few tens of binary separations \citep{Chiang2004,Lodato2013}.  Since a typical protoplanetary disc extends to around $100\,\rm au$ \citep[e.g.][]{Williams2011}, this may be hundreds of binary separations for a close binary. For wider binaries of order a hundred au, a circumbinary disc is unlikely to extend more than a factor of a few binary separations. In this Section we consider the effect of increasing the relative size of the disc. 

Fig.~\ref{size} shows the evolution for discs with larger initial outer
 radii than our standard initial value, $R_{\rm out}=5 a_{\rm b}$. The initial outer 
radii are $10\,a_{\rm b}$ (left), $20 \,a_{\rm b}$ (middle) and $50\,a_{\rm b}$ (right).  Note that the simulations are resolved with $\left< h \right> /H <1$ for radii $R>2\,a_{\rm b}$ for the duration of the simulations. In
each case, the solid line shows the evolution of the inner disc and the
dashed line shows the evolution at the initial outer radius of the
disc. The simulations shown are run4, run5 and run6 in
Table~\ref{tab}.  Since the oscillation period for the test particles
increases with distance from the binary, the larger the initial disc
size, the longer the timescale for the oscillations. Also, the larger
the disc, the more rapidly damped are the oscillations. 

The red line in each lower plot is  the azimuthal angle of the eccentricity vector of the binary.
The eccentricity vector of the binary is denoted by $\bm{e_{\rm b}}=(e_{{\rm b}x},e_{{\rm b}y},e_{{\rm b}z})$. 
Initially, this vector lies along the $y$-direction (by the definition of the $y$--direction), that is, $\bm{e_{\rm b}}(t=0)=(0,0.5,0)$.
In the case of a disc with zero mass, this vector remains constant in time. However, a disc with non--negligible angular momentum causes the eccentricity vector of the binary to precess. We define the azimuthal angle of the eccentricity vector as 
\begin{equation}
\phi_{\rm b}=\tan^{-1}\left(\frac{e_{{\rm b}y}}{e_{{\rm b}x}}\right).
\label{ephi}
\end{equation}
Recall that the precession angles are defined relative to the initial orientation of the binary.
The eccentricity vector 
changes somewhat in time in response to changes in the orientation of the disc.
Notice in the left two panels, the precession angles tend towards equalising over time, indicative of near polar alignment
of the disc with the binary. The large disc plotted in the right panel also shows  signs of evolution towards the polar state
at both $R= 3 a_{\rm b}$ and $R = 50 a_{\rm b}$, although the evolution is much slower at the larger radius.  

The solid and dashed lines in Fig.~\ref{size} that respectively represent the inner and outer parts of the disc increase in separation  with increasing initial outer disc radius. 
As seen in Fig.~\ref{size}, the narrowest disc behaves like a rigid (flat) body, since the evolution is nearly independent of radius
within the disc. 
But as the initial outer disc
radius increases, the disc becomes more warped. The warping occurs because the communication timescale
across the disc becomes longer than the precession
timescale.  We can again investigate this result using equation (34) of LM18.
We obtain $R_{\rm break} \simeq 2.1 a_{\rm b},  2.6 a_{\rm b},$ and $3.4 a_{\rm b}$
for the models with $R_{\rm out}= 10 a_{\rm b}, 20 a_{\rm b},$ and $50 a_{\rm b}$, respectively.
These results are consistent with the strong breaking at $R > 3 a_{\rm b}$ seen in the upper right plot
for $R_{\rm out}= 50 a_{\rm b}$. 

We now examine the largest initial disc outer radius case in more detail. Fig.~\ref{warp} shows the disc structure (surface density, inclination and longitude of ascending node as a function of radius)
at four different times for the model with the larger initial outer disc
radius of $50 \,a_{\rm b}$. Initially the disc is inclined by
$60^\circ$ across all radii. This disc shows a warp in the
inclination and a twist in the precession angle that moves outwards in
time. The warp propagation timescale is about 10\% of the sound speed
in the disc. However, we do note that breaking is observed only for high resolution simulations and for lower resolution the disc may look warped \citep{Nealon2015}. It is possible that with higher resolution this disc would be seen to be broken rather than just warped. In this case, rings may break off from the inner parts of the disc and move to polar alignment in radially discrete sections. Over a longer timescale we expect that the disc will
continue to evolve towards polar alignment. Thus, the disc
size does not affect the outcome, at least for a low mass disc that lasts sufficiently long. The disc size affects only the timescale on which the
alignment occurs and the level of warping in the disc during its evolution towards polar. This result may have implications for planet formation around binaries that we discuss further in Section~\ref{discussion}.

\subsection{Effect of the inclination}
\begin{figure*}
\centering
\includegraphics[width=8cm]{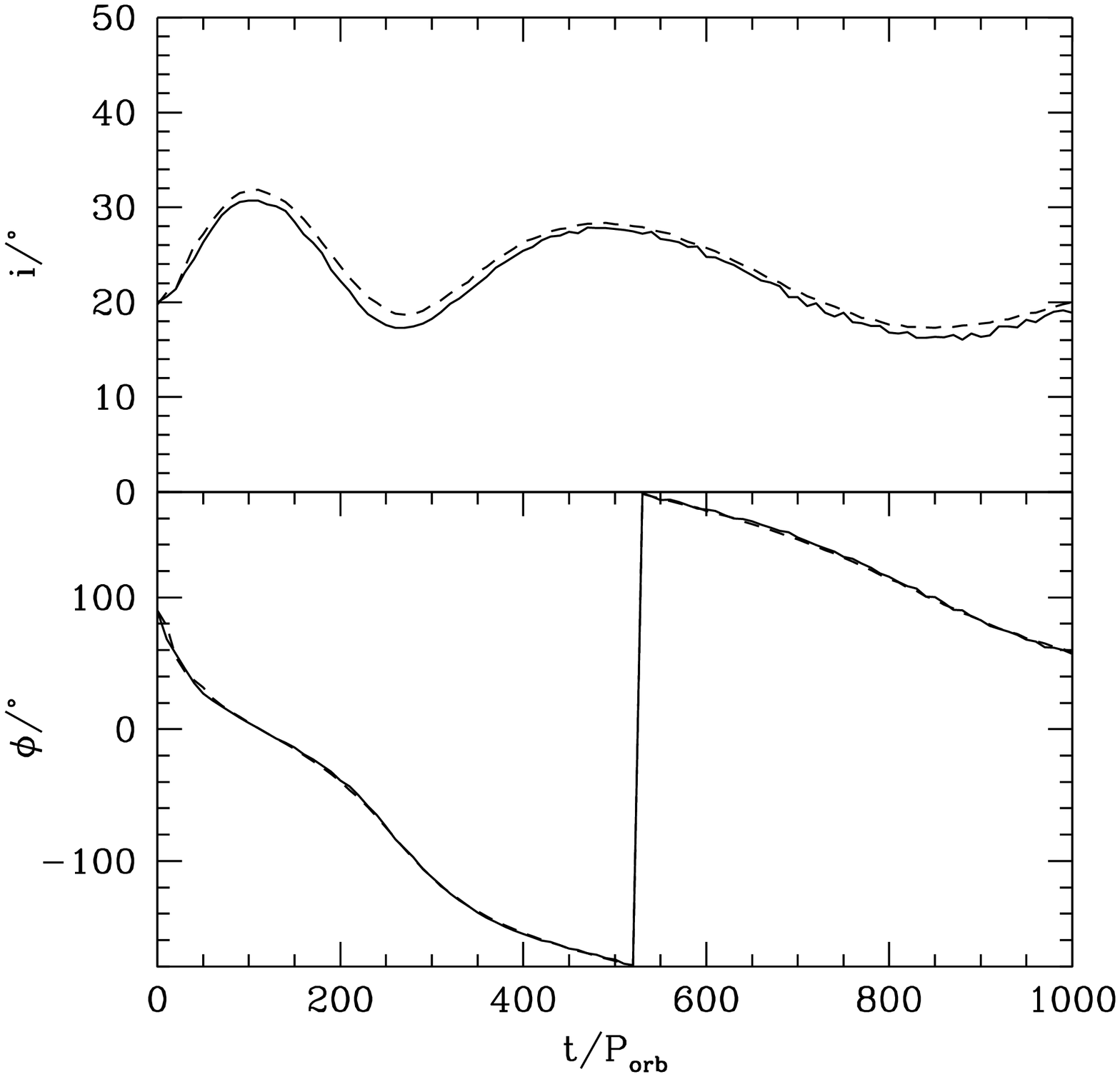}
\includegraphics[width=8cm]{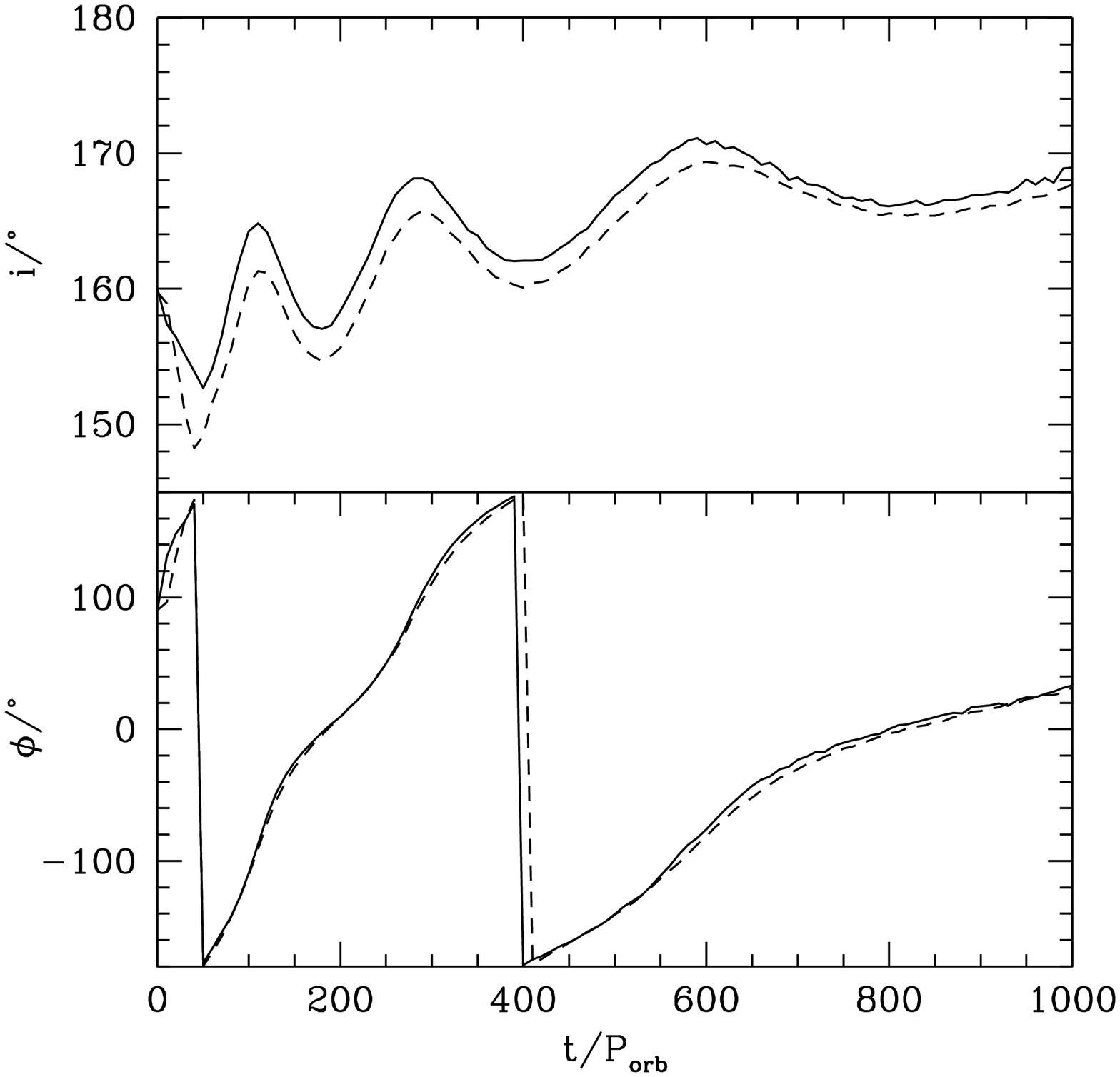}
\caption{Same as Fig.~\ref{paper1} with $3 \times 10^5$ particles, except for the initial disc
  inclination of $i=20^\circ$ (left, run7) and $i=160^\circ$
  (right, run10)  at $R=3
  \,a_{\rm b}$ (solid line) and $R=5 \, a_{\rm b}$ (dashed line). }
\label{inc}
\end{figure*}

Fig.~\ref{inc} shows the disc evolution with a lower inclination of $20^\circ$ in the left panel (run7). As predicted by test particle orbits, the disc displays a circulating behaviour for the longitude of the ascending node and oscillatory behaviour of inclination, whilst slowly aligning the binary orbital plane. There is a critical inclination above which the disc is in the librating state. For our standard disc parameters, a disc that is initially at an inclination of $40^\circ$ is in the circulating state (run8), while a disc that begins at an inclination of $50^\circ$ is in the librating state (run9). Likely due to the mass of the disc, the critical inclination is slightly higher than that predicted by test particle orbits for a binary eccentricity of $0.5$ that is just under $40^\circ$, as
given by equation (2) of \cite{Doolin2011}.

The right panel of Fig.~\ref{inc} shows the
evolution of a disc that begins close to counter alignment at
$i=160^\circ$. The disc evolves towards counter alignment. The tilt
oscillation and precession timescales are shorter in this case likely because
the disc extends closer
to the binary \citep{Nixon2015}.

\subsection{Effect of the binary mass ratio}

\begin{figure*}
\centering
\includegraphics[width=8cm]{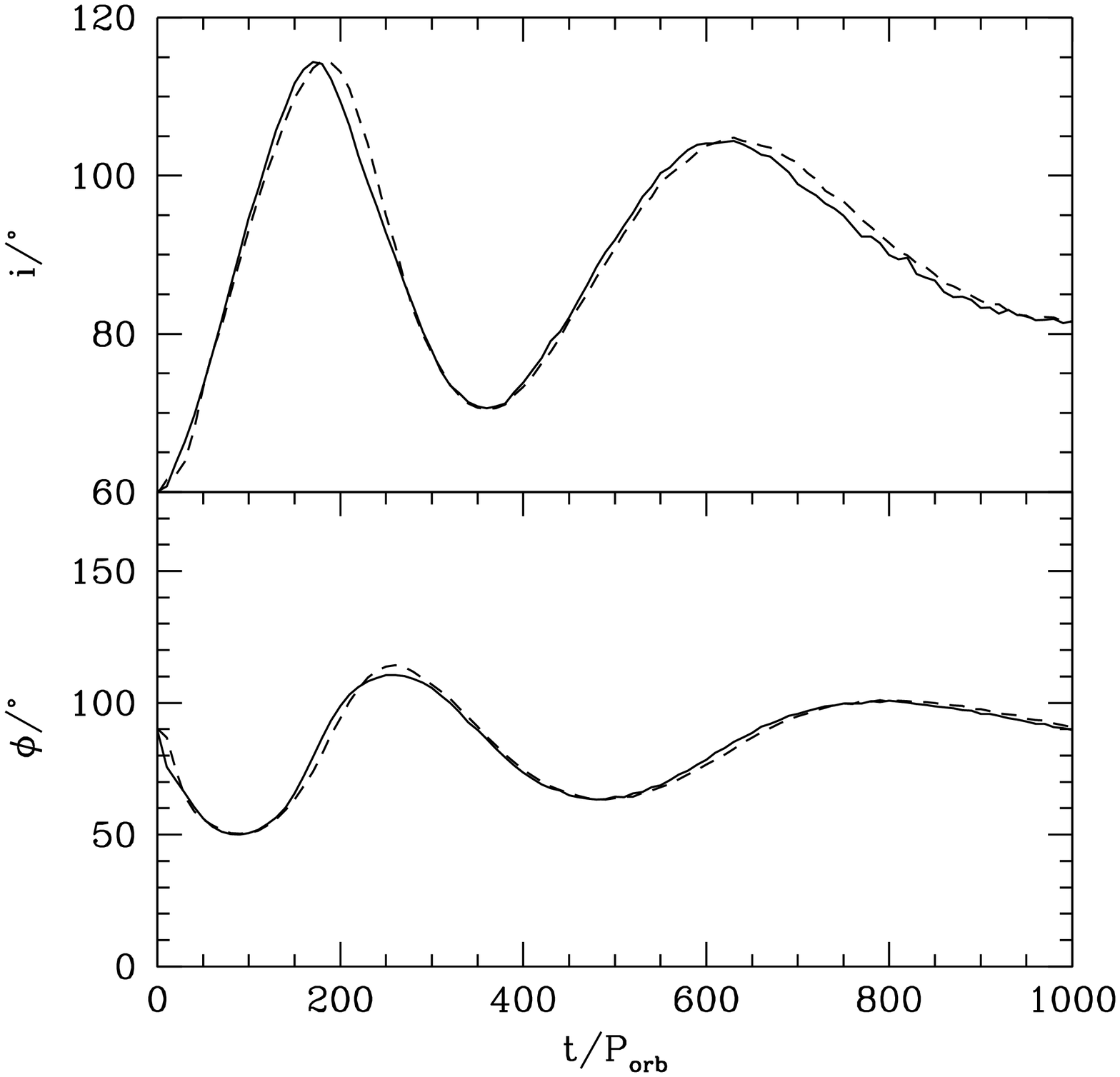}
\includegraphics[width=8cm]{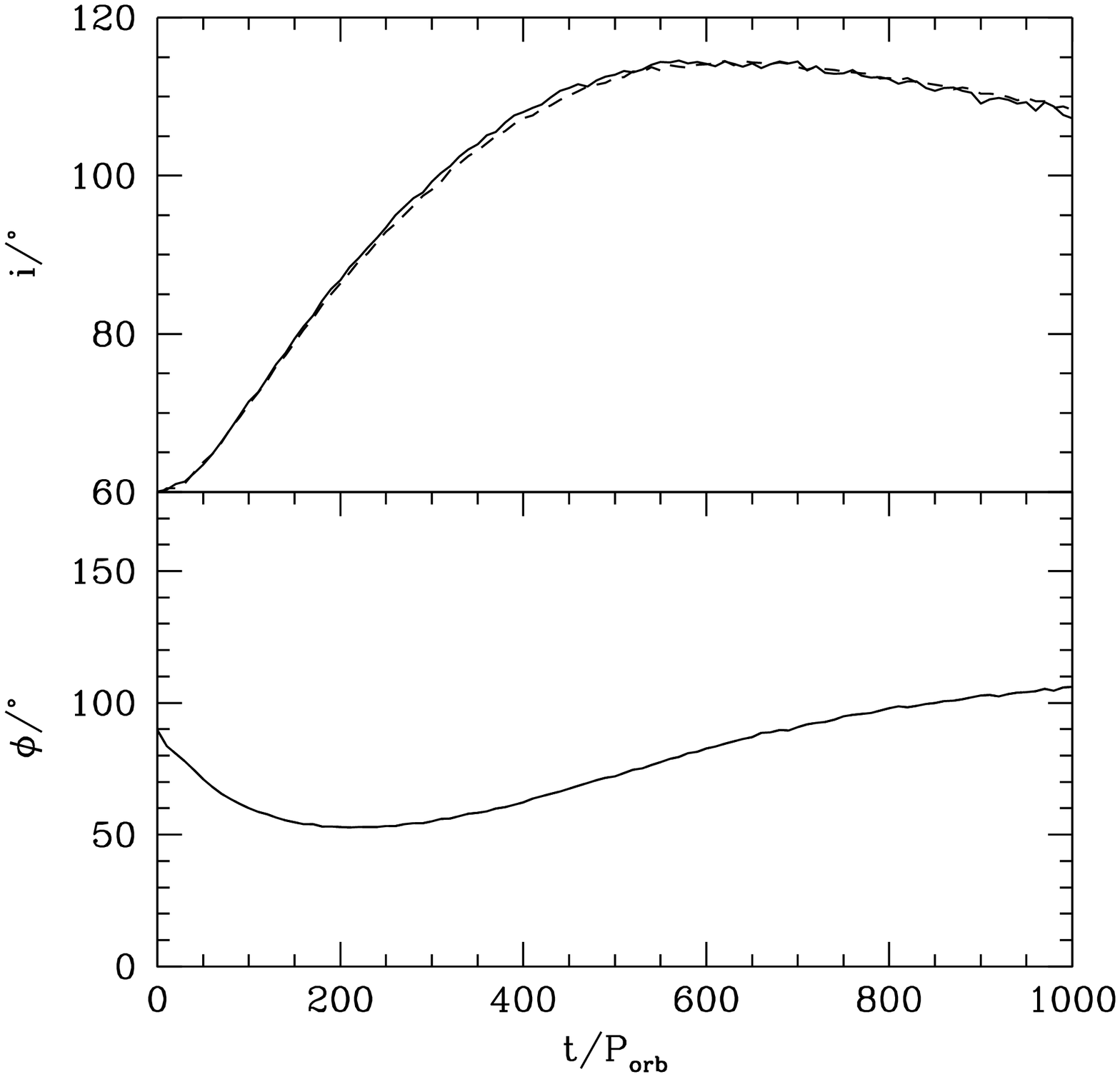}
\caption{Same as Fig.~\ref{paper1} with $3 \times 10^5$ particles, but with a binary mass ratio of $M_2/M_1=0.3/0.7$ (left, run11) and $M_2/M_1=0.1/0.9$ (right, run12)  at $R=3
  \,a_{\rm b}$ (solid lines) and $R=5 \, a_{\rm b}$ (dashed lines). The phase angles at the two different radii are indistinguishable in the lower right hand plot.
  }
\label{q}
\end{figure*}

Fig.~\ref{q} shows the effect of changing the binary mass ratio. The left panel
shows $q=0.3/0.7$ (run11) and the right panel $q=0.1/0.9$ (run12). The behaviour is
qualitatively the same as for our standard parameters. The disc evolves
towards polar alignment.  That is, the angular momentum vector of the disc
aligns with the eccentricity vector of the binary. However, the
decrease of binary mass ratio causes an increase the oscillation period (a decrease
in the precessional frequency), as indicated by equation (16) of LM18. 
This equation suggests an increase in precession period  by a factor of 1.2 and 2.8, for $q =0.3/0.7$ and 0.1/0.9,
respectively for a disc of fixed structure.  In the case of $q=0.1/0.9$, the time coverage of the simulation is not adequate
to obtain a precession period, but it is clear that the tilt oscillation period has increased substantially.

For the case of $q=0.3/0.7$, the
alignment timescale is slightly increased compared to the standard model with mass ratio unity.
This increase occurs because the disc behaves more rigidly for
lower mass ratio, since the precession timescale is longer, as 
as indicated by equation (30) of LM18. 
This equation suggests an increase in tilt evolution timescale for fixed disc structure by a factor of 1.4 and 7.8, for $q =0.3/0.7$ and 0.1/0.9,
respectively.

\subsection{Effect of the binary eccentricity}

\begin{figure*}
\centering
\includegraphics[width=8cm]{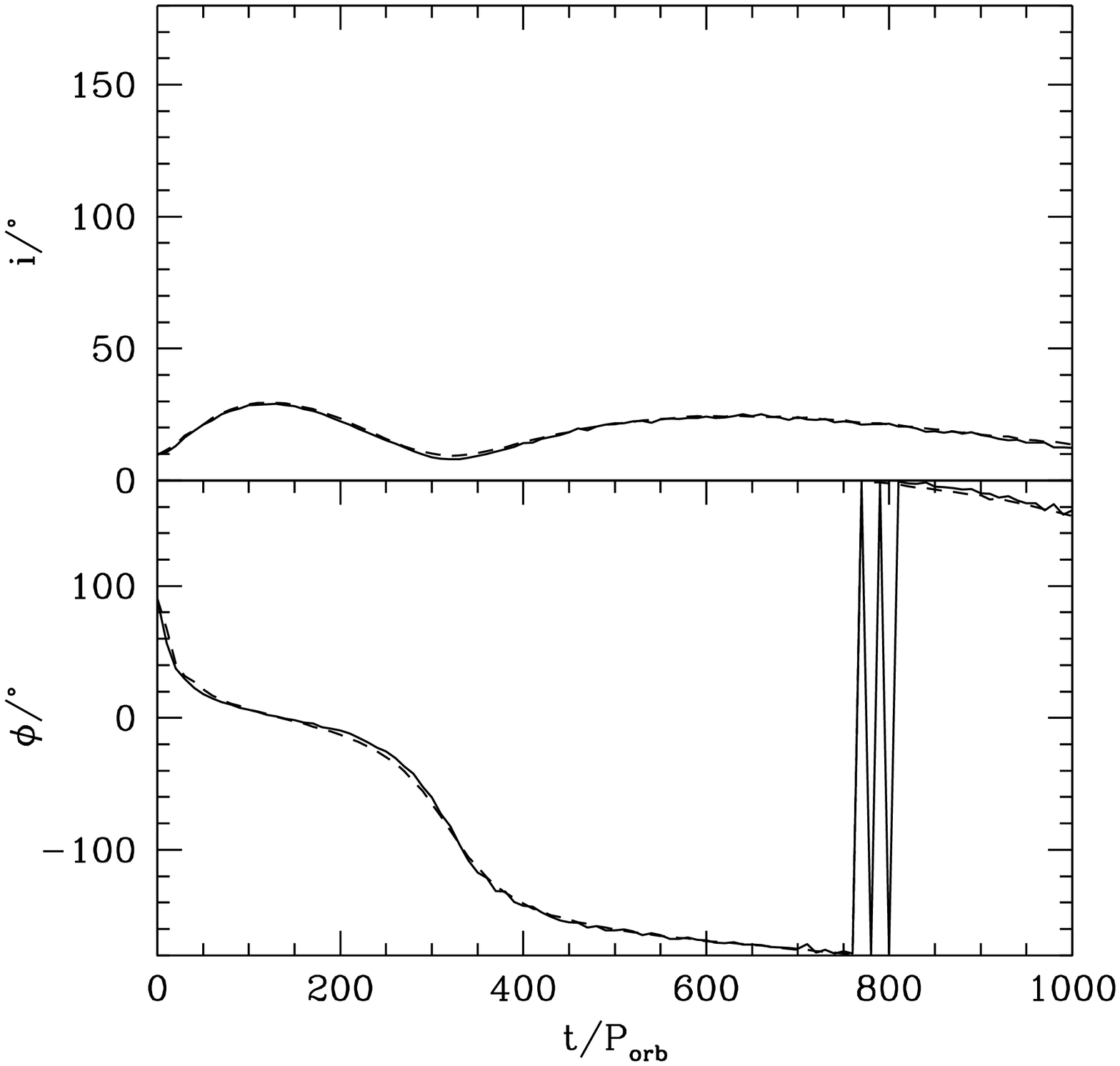}
\includegraphics[width=8cm]{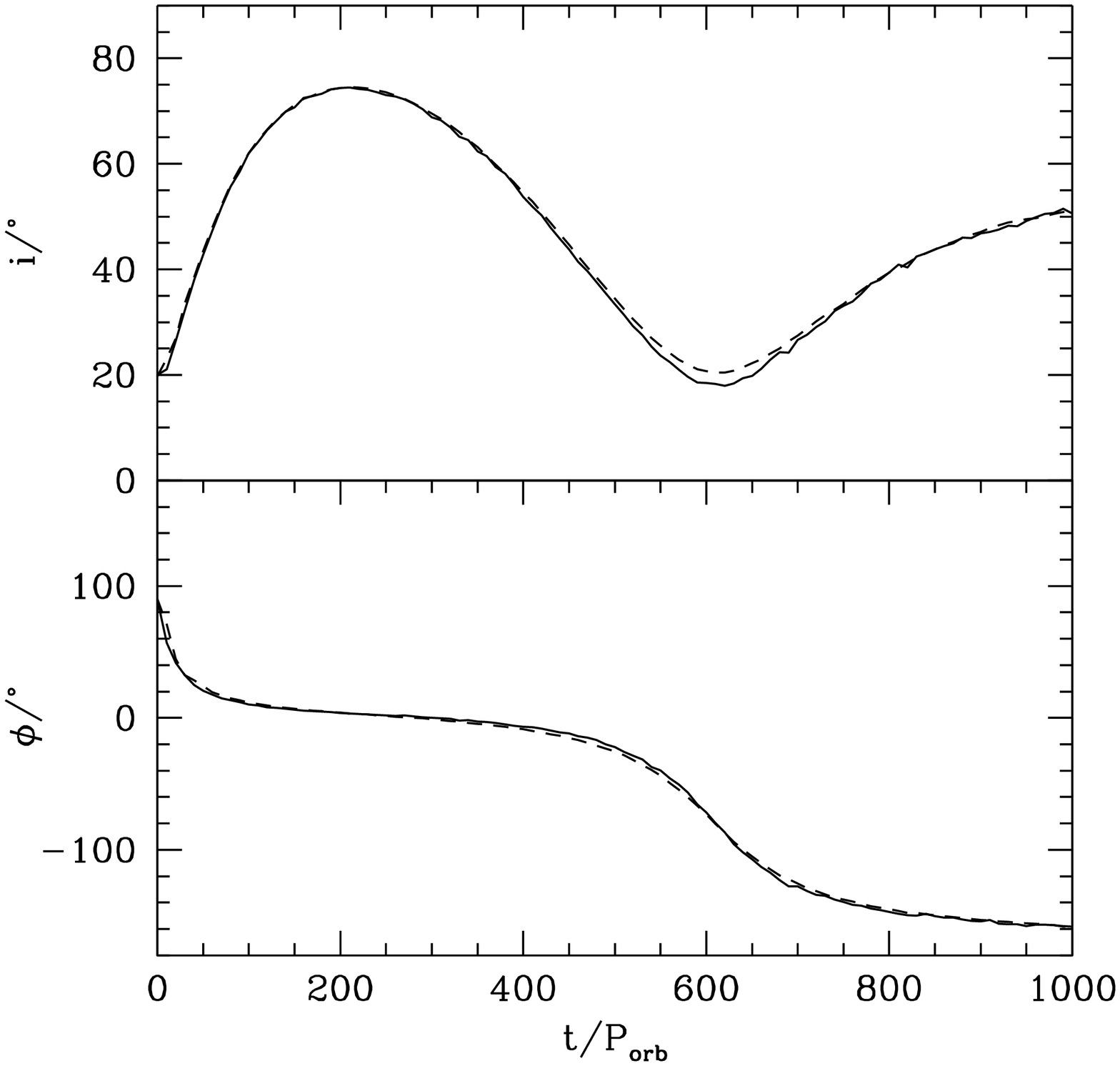}
\includegraphics[width=8cm]{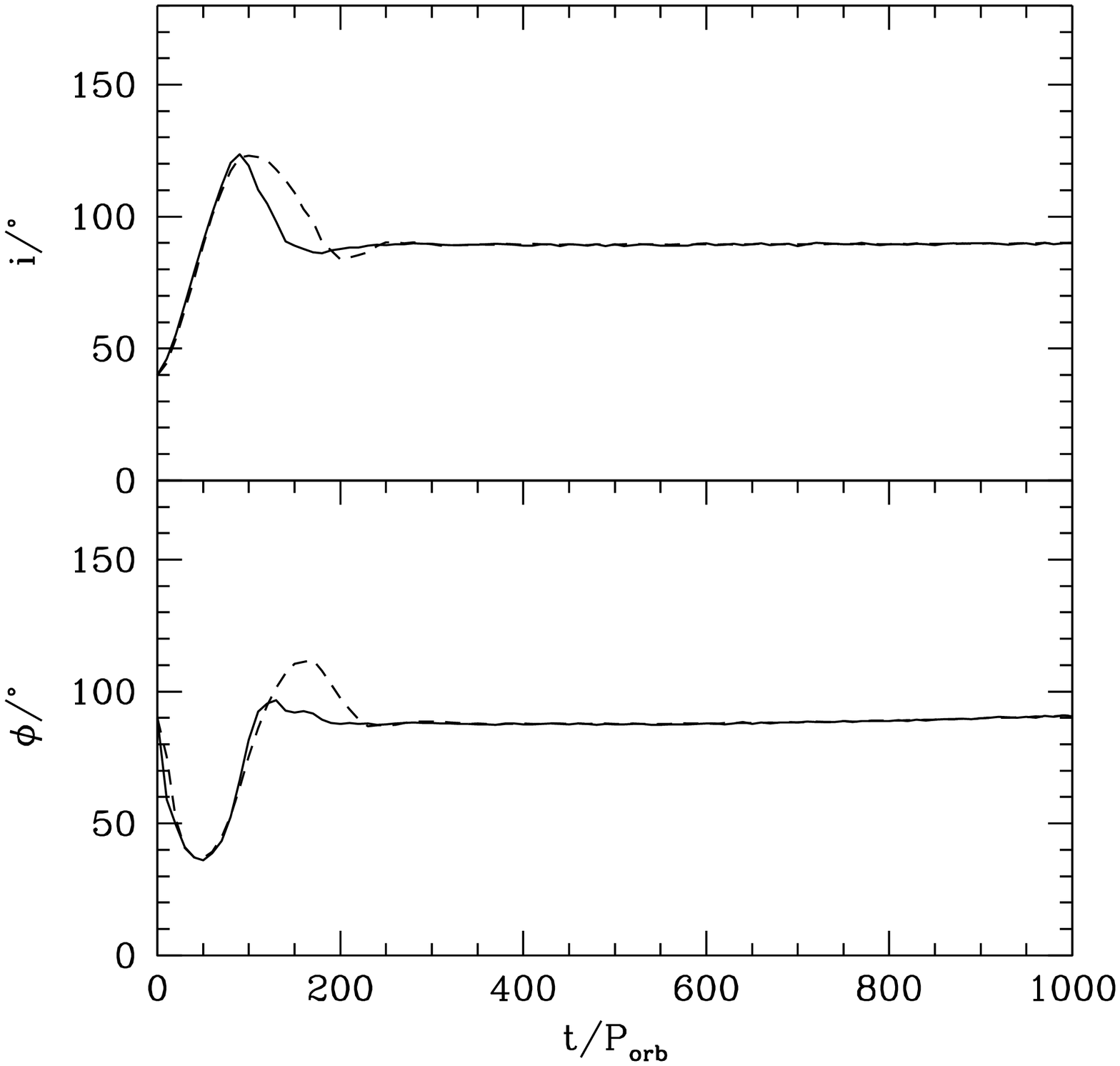}
\includegraphics[width=8cm]{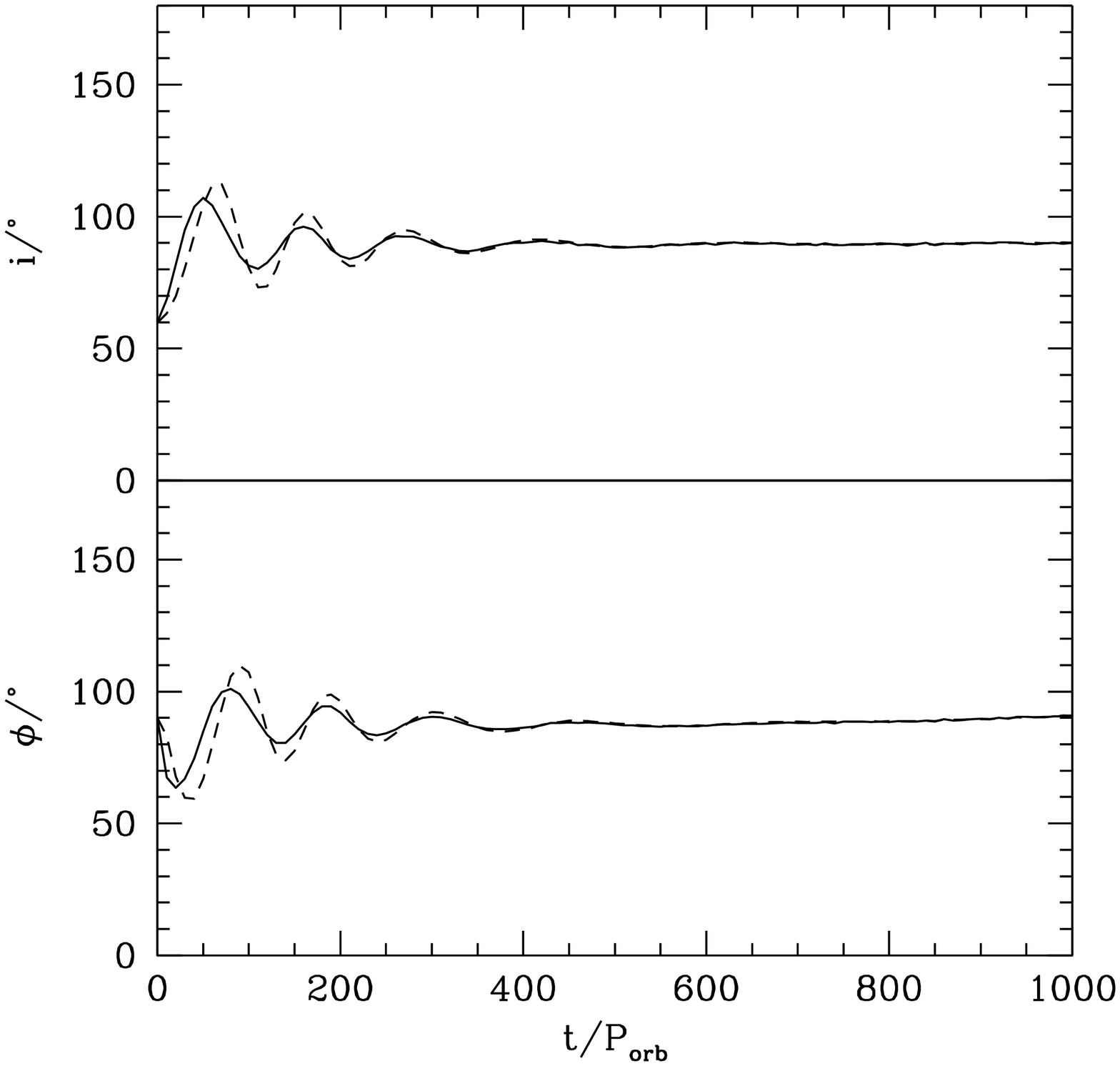}
\caption{Effect of a higher binary eccentricity of $e_{\rm b}=0.8$ for initial
  inclination of $10^\circ$ (run13, top left panel), $20^\circ$ (run14, top right
  panel), $40^\circ$ (run16, bottom left panel) and $60^\circ$ (run17, bottom right
  panel) at $R=3
  \,a_{\rm b}$ (solid line) and $R=5 \, a_{\rm b}$ (dashed line). }
\label{kh15d}
\end{figure*}

In Fig.~\ref{kh15d} we now consider a model with a higher binary eccentricity $e=0.8$ for
varying initial inclination. This is applicable to the observed circumbinary disc in KH~15D. We consider various inclinations as shown in run13, run14, run15, run16 and run17 in Table~\ref{tab}.  We find that the nodal libration mechanism operates for inclination $30^\circ$ but not for $20^\circ$. The critical angle for libration of test particles is given by equation (2) of \cite{Doolin2011}, which
suggests $i_{\rm crit}= 18.5^{\circ}$, a somewhat smaller angle than we find for the disc case we have simulated.

For  such a high eccentricity binary, tilt oscillations occur no matter what the initial inclination of the disc.  Even when the disc is precessing about the binary angular momentum (the circulating solution) with small initial tilt, there can be significant tilt oscillations. The oscillation period is much shorter when the oscillations are librating rather than circulating. The amplitude of the oscillations is much larger for circulating discs rather than librating. For example, the disc that begins at an inclination of $20^\circ$ oscillates up to an inclination of greater than $70^\circ$ before returning to smaller values. Thus, no matter what the inclination of the disc in KH~15D, we expect that the inclination will not damp monotonically down to the binary orbital plane. Furthermore, the timescale for the alignment is much shorter for a librating disc that aligns to polar compared to a circulating disc that eventually will align with the binary orbital plane. 
We plan to explore these effects further.

\section{Comparison of simulations with linear theory}
\label{linth}

\begin{figure*}
\centering
\includegraphics[width=8cm]{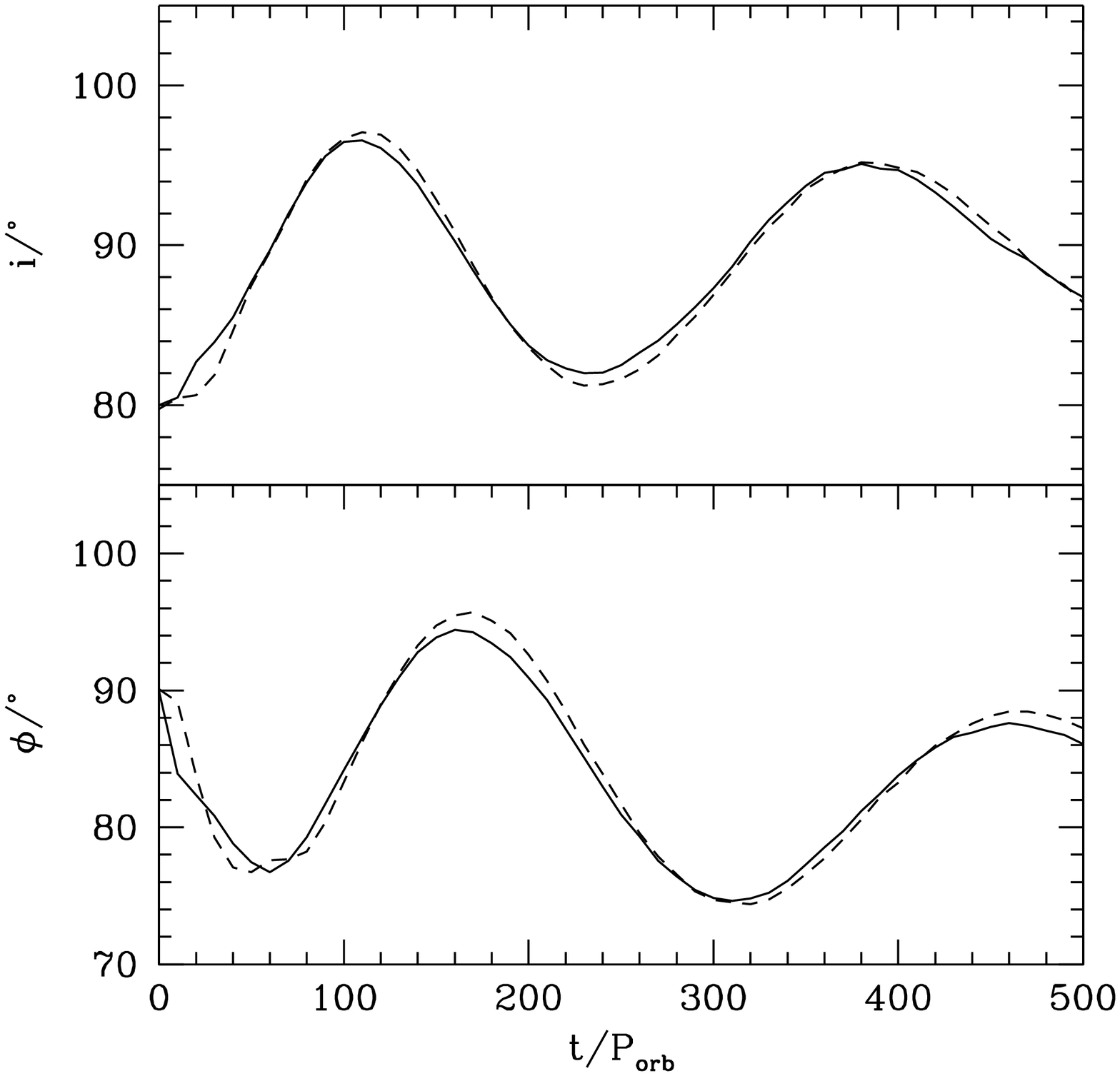}
\includegraphics[width=8cm]{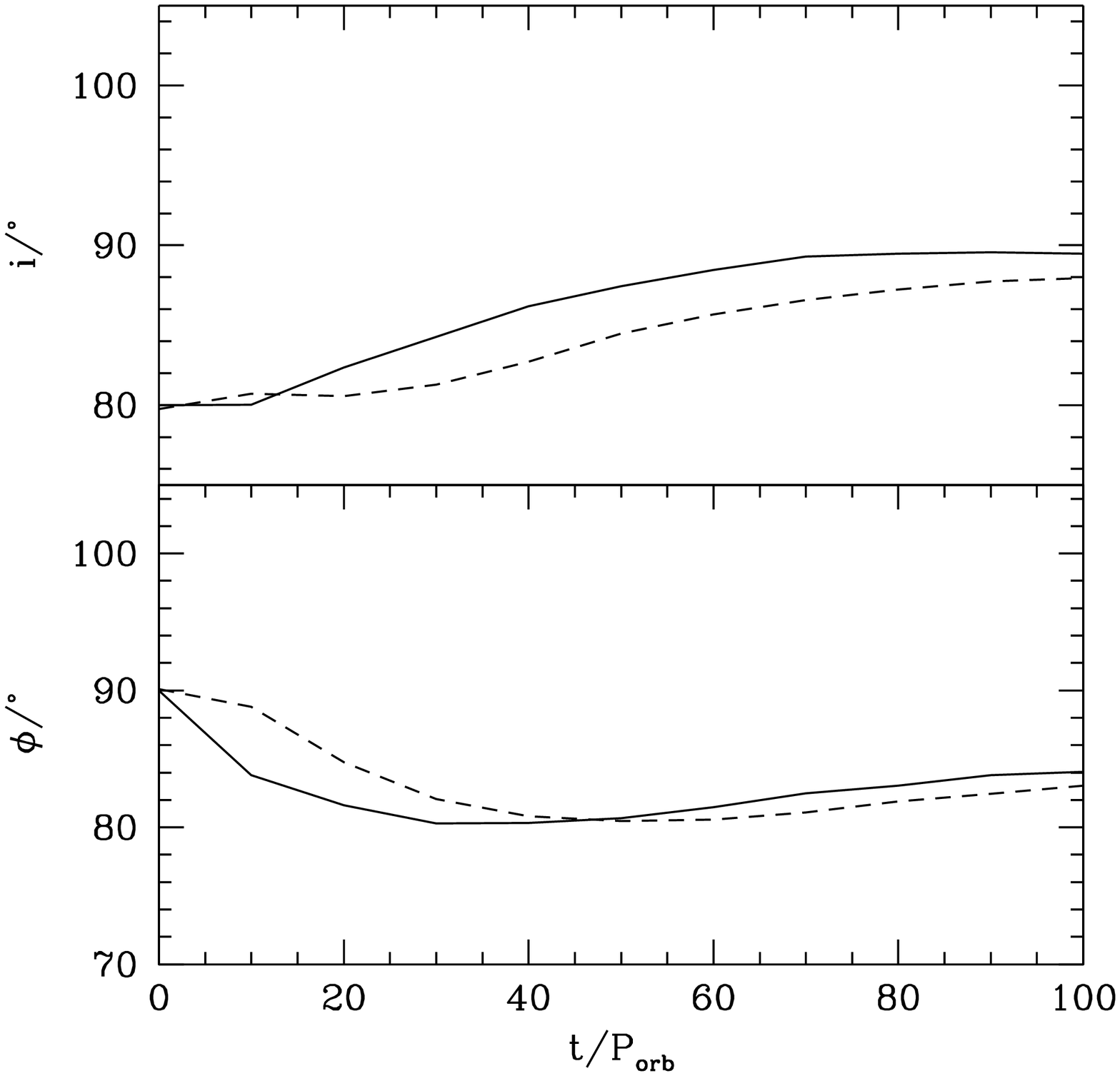}
\caption{A disc that begins almost polar with an inclination of $80^\circ$ to the binary orbital plane with $\alpha=0.01$ (left, run18) and $\alpha=0.1$ (right, run19). We plot the inclination with respect to the (initial) binary orbital plane (upper panels) and the longitude of the ascending node (lower panels)
at $R=3\, a_{\rm b}$ (solid lines) and $R=5\,a_{\rm b}$ (dashed lines).  }
\label{nearlypolar}
\end{figure*}

\subsection{Simulations}
\label{sec:sim}
We compare some simulation results to the linear theory of warped discs. Linear theory for a polar disc is most applicable
to a disc whose tilt relative to the polar state is small (LM18).  For this purpose, 
simulation run18 involves a small initial tilt away from polar, $10^{\circ}$.
Fig.~\ref{nearlypolar} shows the evolution of discs with initial misalignment angle of $i=80^\circ$ and
 $\alpha=0.01$ (left) and $\alpha=0.1$ (right). 
 Each panel  shows the inclination of the disc with respect to the binary orbital plane 
 and the longitude of the ascending node at $R=3 a_{\rm b}$. The disc is initially truncated at 
 $R_{\rm in}=2\,a_{\rm b}$, but quickly spreads inwards. 

For the $\alpha=0.01$ case, 
we estimate the disc precession period as twice the time between  the first peak and subsequent trough in the lower left
panel of Fig.~\ref{nearlypolar}.  We estimate this precession period as
\begin{equation}
P_{\rm p} \simeq 290 P_{\rm orb}.
\label{preceper}
\end{equation}

Using the upper left panel of Fig.~\ref{nearlypolar}, we determine the times of the local extrema of inclination 
relative to the binary orbital plane for the $\alpha=0.01$ case
to be $t_1=106\, P_{\rm orb}$ (local maximum),  $t_2=234 \,P_{\rm orb}$ (local minimum), and $t_3=380 \,P_{\rm orb}$ (local maximum). 
The tilt decay timescale towards the polar orientation is then estimated as
 \begin{equation}
\tau =  \frac{t_3-t_1}{\log{ [ (i_1- i_{\rm m} )/(i_3 - i_{\rm m} )]}} = 1.25\times10^3 P_{\rm orb},
\label{taudecay}
\end{equation}
where $i_k=i(t_k)$ for $k=1,2,3$ and $i_{\rm m}= (i_1+2 i_2 + i_3)/4$
is an estimate of  the midpoint of the inclination oscillations. We expect that $i_{\rm m}=90^{\circ}$
for a massless disc, based on the behaviour of test particles. 
But owing to the nonzero disc mass, there could be departures from that
angle, as is found for the case of particles of nonzero mass \citep{Farago2010}. 
We find that $i_{\rm m} = 88.9^{\circ}$.
The figure also shows that the evolution at two different disc radii,
$3 a_{\rm b}$ and $5 a_{\rm b}$ is quite similar, indicating that the disc is quite
flat over these radii. 

For the $\alpha=0.1$ case, the evolution is much more rapid, after 
a timescale of $ \sim 100 P_{\rm orb}$ the inclination settles to nearly $90^{\circ}$.
From the lower right panel, we see that the disc also quickly settles into polar alignment 
with $\phi \simeq 90^{\circ}$. Evidently, because the settling timescale is short compared
to the disc precession period, the disc does not undergo nodal and tilt oscillations, as  found 
in the $\alpha=0.01$ case. This result is not surprising because the tilt evolution rate  increases with the disc viscosity (LM18).

\subsection{Linear model}

The linear theory of warped  discs was developed to describe the evolution of gaseous discs
that are nearly flat \citep{Papaloizou1995, Lubow2000}. The theory has been recently applied to highly inclined discs
around eccentric binaries  \citep[LM18,][]{Zanazzi2018}.  The linear model requires an assumed density profile.
 We apply a density profile obtained from the simulation run18.
 We apply linear theory to derive the predicted disc precession period and tilt decay timescale towards
the polar configuration. We then compare these results to those obtained from the simulation for run18, as decribed in Section \ref{sec:sim}.

In the linear theory, the disc is taken to consist of circular rings at each disc
radius $R$. 
The rings are subject to an external torque due to the binary, as well as an internal torque due to interactions between the rings.
Each ring is characterised by a vector that is parallel to its angular momentum vector.
For small tilts relative to some reference plane, we solve the
equations in given in \cite{Lubow2000} for the evolution of the "horizontal" projection (along the reference plane) of that   
vector denoted by $\bm{\ell}(R, t)$, a two-dimensional tilt vector. This tilt vector has the property that $|\bm{\ell}(R, t)|=\ell(R,t)=i(R,t)$, the tilt angle, for small tilts.
In the current case of a nearly polar disc, the reference  plane lies perpendicular to the binary eccentricity vector. 
The rings interact through pressure and viscous forces. Together with boundary conditions, the linear equations
 governing the evolution of $\bm{\ell}$ at each
radius are solved for eigenmodes of the form $\bm{\ell}(R) \exp{(i \omega t)}$, where the real part of eigenvalue $\omega$ 
determines the precession rate and the imaginary part determines the tilt growth/decay rate. There are infinitely many
modes possible for each disc model. The fundamental (lowest order) mode is the longest lived and is the relevant one
for the study of a warped disc. 
LM18 determined the precession rates and inclination evolution rates of warped discs towards the polar configuration  by means of linear
modes for discs with power law density profiles in radius. 

\begin{figure}
\centering
\includegraphics[width=6 cm]{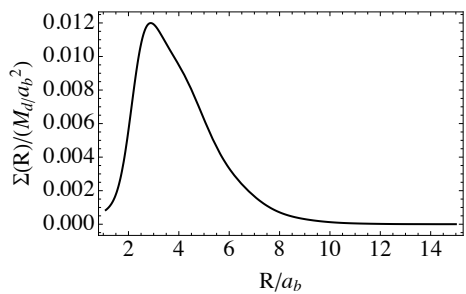}
\caption{Disc density profile for the  model in the left panels of Fig.~\ref{nearlypolar}  model (run18 in Table \ref{tab}) at a time of $500 P_{\rm orb}$.}
\label{simdens500}
\end{figure}

We apply the same procedure here, but for the density profile of run18 at a time of $500 P_{\rm orb}$ that is plotted in
Fig.~\ref{simdens500}.  The density profile drops to small values at $R=1.1 a_b$ that we regard as the disc inner radius.  For such a nearly polar disc, the disc inner truncation may be 
due to the  so-called $l=1, m=2$ resonance that occurs at $\Omega(R) = \Omega_{\rm b}/3$ or $R \simeq 2.1 a_{\rm b}$. From  
the upper left panel of Fig. 8 in LM18, we see that this resonance is capable of truncating the standard model disc ($\alpha = 0.01 <
\alpha_{\rm crit} \simeq 0.02$  for $e_{\rm b}=0.5$ in the figure).
It is likely that mass leaks inward of $1.1 a_{\rm b}$, as is known for the coplanar case \citep{Artymowicz1996}.
But such flows are rapid, involve small densities, and have little influence on the disc precession or tilt evolution rates.

\begin{figure}
\centering
\includegraphics[width=6cm]{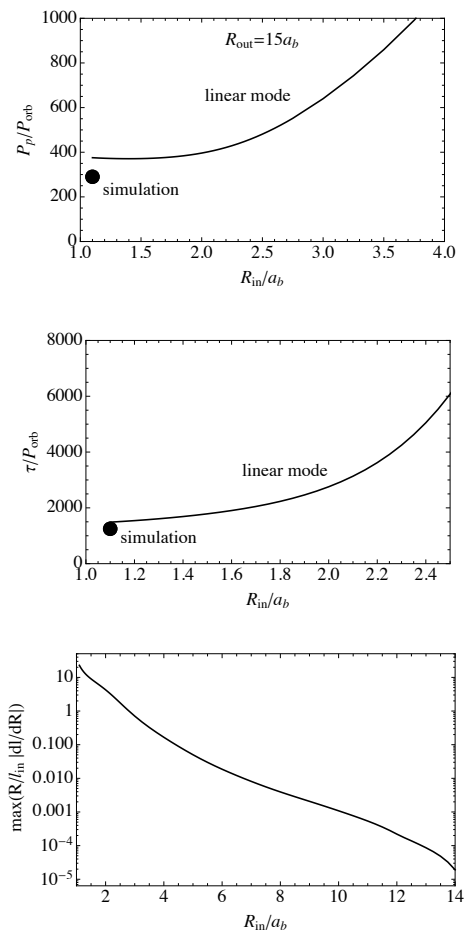}
\caption{Properties of linear modes for the standard disc model at a time of $500 P_{\rm orb}$ with a disc outer radius at $R_{\rm out}=15 a_{\rm b}$
and the various values of the disc inner radius plotted on the horizontal axis. From top to bottom we plot on the line the disc precession period, tilt decay timescale towards polar orientation, and maximum normalised warp within the disc. The large dots plot the corresponding simulation results.  }
\label{modes15}
\end{figure}

The outer part of the disc has an extended tail. 
The disc  is taken to have an outer radius of $R_{\rm out}=15 a_{\rm b}$ that includes essentially the outermost extent of
the density in Fig.~\ref{simdens500}. We consider a sequence of linear models with varying disc inner radii $R_{\rm in}$
that truncate the density plotted in Fig.~\ref{simdens500}.
For each linear model, we determine the disc precession period $P_p$ and the decay timescale towards the polar orientation $\tau$ from its eigenvalue. 
In Fig.~\ref{modes15} we plot results of the linear mode calculations. 
From the line on the upper panel of the figure we see that the precession period decreases with decreasing disc inner radius, since
the disc experiences a stronger tidal torque from the binary. For $R_{\rm in} \la 2 a_{\rm b}$, the precession period
becomes insensitive to the disc inner radius because the disc density is relatively low. 
In the middle panel, we plot the tilt decay timescale  towards the polar orientation  as a  solid line. This quantity also becomes less sensitive to the disc inner
radius at smaller inner radii, again as expected by the lower density. However, it is more sensitive to $R_{\rm in}$ than
the precession period.  Also plotted as solid dots are the quantities determined by the simulation for run18 in Equations
(\ref{preceper}) and (\ref{taudecay}).  

Although there is approximate agreement between the linear model and simulation results,
we need to check on the validity of the linear approximation. The linear model assumes that the level of warping is small. 
The level of warping is measured by $|R d \bm{\ell}/dR|$ where $\bm{\ell}(R)$ is a disc tilt vector.
For a flat disc, $\bm{\ell}$ is independent of radius. Just how small  $|R d \bm{\ell}/dR|$ needs to be in order to
justify the linear approximation is not well known. \cite{Ogilvie2006} showed that nonlinearlity is important
in affecting the shape of bending waves for   $|R d \bm{\ell}/dR| \sim (H/R)^{1/2} \sim 0.3$.
Also, \cite{Nelson1999} found nonlinearlity in the form of shocks can arise if the warp wavelength is short enough that the shearing motions from different parts cause a collision. This effect typically requires $|R d \bm{\ell}/dR|$ to be of order unity.

We estimate the level of warping in these disc models for various disc inner radii.
In the linear theory as applied in LM18, there is an arbitrary tilt amplitude at the disc
inner edge $\ell_{\rm in}= |\bm{\ell}(R_{\rm in})|$ that is set to unity in carrying out the mode calculations.  
The tilt at each radius scales linearly with $\ell_{\rm in}$. From our knowledge
of $\ell_{\rm in}$ obtained through the simulations, we can scale the values of $\bm{\ell}(R)$ 
obtained from the modal calculation by this value of $\bm{\ell}_{\rm in}$ and calculate $\bm{\ell}(R)$ and the warping 
appropriate to the simulation based on linear theory.

The linear modes calculations for a given disc model determine the normalised warping $|R/\ell_{\rm in} d \bm{\ell}/dR|$.
To estimate the level of normalised disc warping in each disc model having some $R_{\rm in}$,
we calculate the maximum value of this normalised level of warping across all radii in the disc. 
(We note that the warping levels in
Section 4.4 of LP18 are estimated
with a normalisation such that $\ell_{\rm in}=1$. In this paper, we properly normalise these warping levels by $\ell_{\rm in}$.)
The level of warping depends somewhat on the precession phase angle of the disc, since the tilt of the disc varies somewhat with this angle.
For definiteness, we determine the maximum normalised level of warping across all radii in the disc in which the phase at the inner edge of the disc is such that the tilt vector lies in the orbital plane of the binary. 
The bottom panel of Fig.~\ref{modes15} plots the maximum normalised level of warping across the disc as a function
of disc inner radius $R_{\rm in}$.
For this model at this time, the tilt oscillation amplitude of about $6^{\circ}$ and so $\ell_{\rm in}\sim 0.1$.  
It then follows that for $R_{\rm in} \la 2.2 a_{\rm b}$ the linearity condition given by  \cite{Ogilvie2006} becomes violated, that is  
$|R /\ell_{\rm in}\, d \bm{\ell}/dR| \sim (H/R)^{1/2}/\ell_{\rm in} \ga 3$.

\begin{figure}
\centering
\includegraphics[width=6cm]{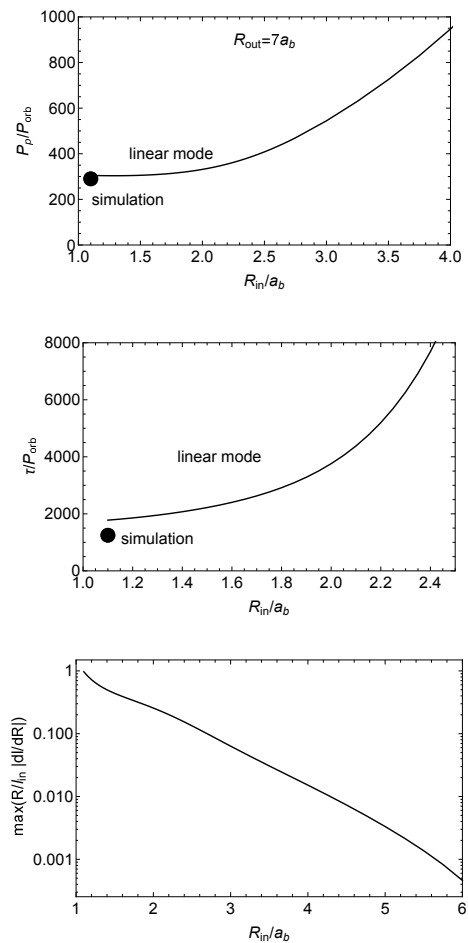}
\caption{
Same as Fig.~\ref{modes15}, except that  the disc outer radius is located at $R_{\rm out}=7 a_{\rm b}$.}
\label{modes7}
\end{figure}

Strong levels of warping occur for the linear modes
in both the inner and outer parts of the disc where the densities are somewhat low, as is seen for example in Fig.~\ref{inc500}.
  To understand
the contributions of these regions, we consider another model that  also uses the density profile from Fig.~\ref{simdens500} in which 
disc  is taken to be truncated at outer radius $R_{\rm out}=7 a_{\rm b}$. About 10\% of the disc mass extends beyond $R > 7 a_{\rm b}$ in Fig.~\ref{simdens500}. 
From the bottom
panel of  Fig.~\ref{modes7}  we see that the level of warping is much smaller than for bottom
panel of  Fig.~\ref{modes15} in the region where $R_{\rm in} < 2 a_{\rm b}$. The linearity constraint discussed above is satisfied for
this truncated disc even for the smallest inner disc radius, $R_{\rm in} = 1.1 a_{\rm b}$.

The results suggest that the agreement between linear theory and simulations
involves substantial contributions in the disc where the conditions for
linearity hold.  The simulation results are consistent with linear theory  to within a factor of 2 or less.
There is also evidence for some nonlinear contributions to the precession period and tilt evolution rate
 due to warping in inner and outer regions of the disc where the density is low. Nonlinear effects are expected to play a more important
 role in discs that are more radially extended or cooler in which case stronger warping occurs in the main body of the disc.

\section{Discussion}
\label{discussion}

We have shown that the nodal libration mechanism for a circumbinary disc around an eccentric binary leads to polar alignment rather than binary angular momentum alignment for a wide range of binary and disc parameters for a low mass disc. If the disc aligns on a timescale that is shorter than lifetime of the disc, then we expect  that any planetary system that forms will be on a polar orbit.  Without further perturbations, it may remain on  a stable polar orbit (see discussion at the end of LM18).
 A planetary system that forms either polar or aligned with the binary orbital plane may remain coplanar, since it does not undergo differential nodal precession.
 
As discussed at the beginning of Section \ref{sec:discsize}, the size of a circumbinary disc relative to the binary separation can have a wide range of values.
 Close binaries may have a circumbinary disc with a large radial extent compared to the binary semi-major axis. The circumbinary disc may then become significantly warped, as shown in Fig.~\ref{warp}, or even broken. Consider an example of an equal mass binary with masses $M_1=M_2=0.5\,\rm M_\odot$, separation $a_{\rm b}=0.25\,\rm au$ and orbital period $0.125 \,\rm yr$. A  disc with an outer radius of $100\,\rm au$ extends to $400\, a_{\rm b}$. This radius is much larger than anything we have simulated in this work. However, extrapolating from the simulations in Fig.~\ref{size}, this disc will likely break. The alignment timescale for the disc increases with increasing disc size. The inner parts of the disc, however, may move towards polar alignment. 
 
 For sufficiently large discs, the alignment timescale of the outer parts of the disc may be longer than the lifetime of the disc. The angular momentum of the outer parts of a circumbinary disc may remain close to its initial value at formation.  In this case, if the inner parts of the disc are  misaligned with respect to the outer parts of the disc during the time of planet formation, then  a non--coplanar planetary system may form with some of the planets residing on nonpolar,  noncoplanar orbits with respect to the binary. 
If the interactions between the
planets are weak, then the planets will precess with a period that depends only on their distance from the centre of mass of the binary. Differential precession will lead to the planetary system being distributed over a large volume \citep[see also][]{Nesvold2016}. 
 
  Wide binaries likely have a circumbinary disc with a more limited radial extent compared to the binary semi-major axis. Consider an example of a binary with a separation of $a_{\rm b}=100\,\rm au$. A disc with initial outer radius $\sim 500 \rm{au}$ aligns to polar on a timescale of around a few 1000$P_{\rm orb}$ which is of the order of $10^6\,\rm yr$. 
 Gas giant planets must form in less than the lifetime of the disc of a few Myr, while terrestrial planet formation may only occur after gas disc dispersal \citep{Armitage}. 
 The alignment of any planets formed in discs around wide binaries depends sensitively on the evolution of the disc. The planets may be found in coplanar, polar, or misaligned orbits depending on the relative timescales for planet formation and disc alignment.  The binary 99 Herculis has a semi-major axis $a_{\rm b}=16.5\,\rm au$. Its polar debris disc is observed to have a detectable radial extent of about $120\,\rm au$
 that is only about $7\,a_{\rm b}$. Thus, the gas disc was able to align to polar during its lifetime, leaving the debris disc in a stable polar orbit.

In this work we have limited considerations to a small initial disc mass, 0.1\% of
the binary mass. A higher initial mass and angular momentum disc than we have considered would affect the evolution of the binary orbit more significantly, leading to additional effects. We will investigate these effects in a future paper.

\section{Conclusions}
\label{conc}

In this work we have investigated the conditions under which a low mass fluid disc around an eccentric
binary can become aligned perpendicular to the binary orbital plane,
 as first described by \cite{Martin2017}. A sufficiently inclined disc with respect to the binary orbital plane can undergo
oscillations of  tilt and librations of the longitude of ascending node, similar to
test particle orbits \citep{Verrier2009,Farago2010,Doolin2011}. However, for the case of a disc, dissipation can
lead to polar (perpendicular) alignment of the disc. We have considered a range of
disc and binary parameters and found that the mechanism operates over
a wide range of parameters for a low mass disc. The simulation results broadly agree with the expectations of linear theory
(see Fig.~\ref{modes15}). But, nonlinear effects likely play a significant role in the inner and outer parts of the disc. 

The properties of the disc oscillations are similar to the
properties of test particle oscillations. However, the evolution to the polar
state is a distinctive property of discs. Discs with a low temperature are
unable to hold themselves together during the oscillations and this
can lead to disc breaking (see Fig.~\ref{a}).  
 For discs with a large
radial extent, the disc communication timescale between 
the inner and outer regions of the disc  is long.  The oscillations may then lead to a strongly
warped disc (see Fig.~\ref{warp}). Although the polar alignment timescale for a moderately misaligned larger disc is
longer than for a smaller disc, the larger disc also evolves towards polar orientation.  
Discs at low inclination that do not evolve towards
polar states may still undergo significant tilt oscillations (see Fig.~\ref{kh15d}).
Such noncoplanar effects may have important
consequences for planets formed in circumbinary discs.

\section*{Acknowledgments} 
We thank an anonymous referee for useful comments.  We thank Daniel
Price for providing the {\sc phantom} code for SPH simulations and
acknowledge the use of SPLASH \citep{Price2007} for the rendering of
the figures. We acknowledge support from NASA through grant NNX17AB96G.  Computer support was provided by UNLV's National
Supercomputing Center.





\bibliographystyle{mnras} 
\bibliography{martin}

\bsp	
\label{lastpage}
\end{document}